\documentclass[acmsmall, table, nonacm, authorversion]{acmart} 

\newcommand{\bthe}{\begin{theorem}\rm\flushleft}
\newcommand{\ethe}{\rule[-0.5mm]{2mm}{3mm}\end{theorem}}
\newcommand{\context}[1]{}
\newcommand{\progress}[1]{}

\newcommand{\panort}{\textsc{RESOLVE~}}

\newcommand{\mypara}[1]{\vspace*{0.75ex}\noindent{\bf #1}~}

\newcommand{\be}{\begin{itemize}}  
\newcommand{\ee}{\end{itemize}}  
\newcommand{\bn}{\begin{enumerate}}  
\newcommand{\en}{\end{enumerate}}  
\newcommand{\bc}{\begin{center}}  
\newcommand{\ec}{\end{center}}  
\newcommand{\bl}{\begin{flushleft}}  
\newcommand{\el}{\end{flushleft}}  
\newcommand{\bq}{\begin{quote}}  
\newcommand{\eq}{\end{quote}}

\newcommand{\bmp}{\begin{minipage}}  
\newcommand{\emp}{\end{minipage}}

\newtheorem{instance}{Example}
\newcommand{\bexa}{\begin{instance}\rm}
\newcommand{\eexa}{\end{instance}}
\newcommand{\eeeexa}{\end{itemize}\end{instance}}
\newcommand{\eeeeexa}{\rule[-0.1mm]{1.0mm}{3mm}\end{itemize}\end{itemize}\end{example}}

\usepackage{graphics}
\usepackage[export]{adjustbox}
\usepackage[ruled,vlined,linesnumbered]{algorithm2e}
\usepackage{subfig}
\usepackage[utf8]{inputenc}
\usepackage{soul}
\usepackage{pifont}
\usepackage{tikz}
\usepackage{xspace}
\usepackage{multirow}
\usepackage{url}
\usepackage{makecell}
\usepackage{multirow}
\usepackage{verbatim}
\usepackage{lscape}
\usepackage{tablefootnote}
\usepackage{enumitem}
\usepackage{boldline}

\usepackage{algorithm2e}
\SetKwInOut{Parameter}{Parameter}
\usepackage{float}

\definecolor{myredfor3}{HTML}{f4cccc}
\definecolor{myredfor2}{HTML}{ea9999}
\definecolor{myredfor1}{HTML}{e06666}

\definecolor{mycyanfor3}{HTML}{d0e0e3}
\definecolor{mycyanfor4}{HTML}{a2c4c9}
\definecolor{mycyanfor5}{HTML}{45818e}

\definecolor{sharecolor}{HTML}{6d9eed}
\definecolor{notsharecolor}{HTML}{e8b246}

\definecolor{wincolor}{HTML}{43a3a3}
\definecolor{losecolor}{HTML}{cd6dd6}

\setcopyright{none}

\begin{document}
\title{Analyzing Privacy Dynamics within Groups using Gamified Auctions}

\author{H{\"u}seyin Ayd{\i}n*}
\orcid{0000-0002-5746-9702}
\affiliation{\institution{Utrecht University}\country{The Netherlands;}\institution{Middle East Technical University}\country{Turkey}}
\email{h.aydin@uu.nl}

\author{Onuralp Ulusoy*}
\orcid{0000-0003-3737-7947}
\affiliation{\institution{Utrecht University}\country{The Netherlands}}
\email{o.ulusoy@uu.nl}

\author{Ilaria Liccardi}
\orcid{https://orcid.org/0000-0002-3306-505X}
\affiliation{\institution{Massachusetts Institute of Technology}\country{USA}}
\email{ilaria@mit.edu}

\author{P{\i}nar Yolum}
\orcid{0000-0001-7848-1834}
\affiliation{\institution{Utrecht University}\country{The Netherlands}}
\email{p.yolum@uu.nl}

\begin{abstract}
 Online shared content, such as group pictures, often contains information about multiple users. Developing technical solutions to manage the privacy of such "co-owned" content is challenging because each co-owner may have different preferences. Recent technical approaches advocate group-decision mechanisms, including auctions, to decide as how best to resolve these differences. However, it is not clear if users would participate in such mechanisms and if they do, whether they would act altruistically. Understanding the privacy dynamics is crucial to develop effective mechanisms for privacy-respecting collaborative systems. Accordingly, this work develops RESOLVE, a privacy auction game to understand the sharing behavior of users in groups. Our results of users' playing the game show that i) the users’ understanding of individual vs. group privacy differs significantly; ii) often users fight for their preferences even at the cost of others' privacy; and iii) at times users collaborate to fight for the privacy of others.

\end{abstract}

%%
%% The code below is generated by the tool at http://dl.acm.org/ccs.cfm.
%% Please copy and paste the code instead of the example below.
%%
%\begin{CCSXML}
%\end{CCSXML}

%%
%% Keywords. The author(s) should pick words that accurately describe
%% the work being presented. Separate the keywords with commas.

% To generate keywords we need to check --> https://dl.acm.org/ccs



\maketitle

\section{Introduction}
\label{sec:intro}

Privacy is often understood as a right of individuals to choose what information they want to share about themselves~\cite{acquisti2015privacy}. Typical way of treating privacy online is through information consent. In essence, this paradigm puts the individual in charge to declare what information they would like to share, with whom, and for what purposes. General Data Protection Regulation (GDPR) is a prime example of this paradigm, where users are informed about the content that is being accessed about them~\cite{gdpr}. The user then gives an informed consent as to how their personal data will be shared. As a result, with appropriate techniques, it is possible to detect whether personal data have been leaked without the person’s consent or used against their will.

Many times our content is not simply personal data, but pertains to multiple users. Consider a group picture that is posted on social media: There is a single person that uploads the picture, whom might be interpreted as the owner of the content. However, the decision of sharing this content will have privacy implications on the others that are in the picture as well as the uploader. Similarly, in a collaborative editing environment, two individuals might edit the same document and then one of the editors can choose to share the document with others, without the need to consult the other person, again possibly creating privacy violations.
Put simply, these happen because the content that exists on collaborative systems, such as co-edited documents or group pictures, does not have to a single owner, but multiple {\it co-owners}, with possibly different privacy preferences~\cite{Such:2016:PPN:2891451.2821512,hu2012multiparty,sleeper2013post}. Hence, it is not appropriate to assume that the sharing party is the only owner of the content and that her privacy settings should apply to the content.

Protecting the privacy of co-owned content requires a shift from personal privacy to multi-party privacy. Since each user of online content might have different privacy preferences, solutions that aggregate the individual preferences indiscriminately and automatically result in privacy violations of some co-owners~\cite{squicciarini2018multi,Kokciyan:2016:ARP:2937029.2937160}. Ideally, systems should employ group-decision mechanisms that enable each user to express her opinion as to whether a content should be shared and, if so, how. There have been recent technical approaches to create such mechanisms~\cite{Such:2016:PPN:2891451.2821512,Squicciarini:2009:CPM:1526709.1526780,Kokciyan:2017:AAR:3106680.3003434, ulusoy2021panola} without any studies as to whether users would employ and participate in such mechanisms. More importantly, how differently would they behave in a multi-party situation through this type of a mechanism? What would be the factors behind those behaviors? Considering such privacy dynamics of users in groups leads us to these particular research questions:

\begin{enumerate}[label={\textbf{RQ\arabic*:}}]
    \item How do people determine their content sharing preferences in a multi-party situation?
    \item When do people tend to keep their privacy preferences, especially even if they contradict the majority's wishes?
    \item Which circumstances lead people to change their privacy preferences?
\end{enumerate}

This paper addresses these questions by conducting an empirical study on group images.
To realize this study, we have created a novel game \panort that uses a group-decision making system based on auctions, where each player bids on how much they would like to see a {\bf group picture} shared or not shared. The group pictures are created using the participant's own pictures to make them realistic. By creating a systematic quiz for the game, we ensured that the participants understood the rules and workings of the game well. For the purpose of this study, we have one player as participant of the study and three other players that are software agents (configured by us). This enables us to create interesting situations, such as only the participant would not like to share a group picture while the other three would.

Unique to this game is the possibility to revisit privacy choices on the same picture after seeing the preferences of others. By enabling a participant to make a privacy choice before seeing others' choices and afterward, we analyze whether the participants' perception of privacy changes based on the choices of others and whether participants fight for their privacy in such situations. We observe that participants' privacy declarations on individual pictures could change when the individual picture is merged into a group picture. Moreover, some participants actually fight for their privacy even when they know they are in the minority. On the other hand, we also see some participants `switching sides' when their choice is not in the majority, to help others preserve their privacy. These and other observations we describe later are important insights for developing the next-generation software tools to manage the privacy of co-owned data, both on social media as well as in collaborative tools.

Our major contributions are the following:
\begin{itemize}
\item We developed a novel game~\panort. This game is configurable to test user behaviors in various multi-party privacy situations with group pictures of interest.
\item We developed material to teach how to play~\panort as well as a systematic quiz to evaluate and help a user play the game.
\item We have performed a user study with $40$ participants to evaluate privacy dynamics of users while shifting from individual pictures to group contents as well as the effect of knowing the preferences of others.
\end{itemize}

The rest of the paper is organized as follows. Section~\ref{sec:related} summarizes the necessary background information and related work for the study. Section \ref{sec:user_study} presents the user study including all the features of our game \panort and the demographics of the participants. User profiles based on their interactions with the game are provided in Section \ref{sec:profiles_in_multiparty_privacy}. Section \ref{sec:behaviors_in_multiparty_privacy} examines the observed behaviors through the game. Before concluding the paper, the validity and limitations of the study are discussed in Section \ref{sec:validity_and_limitations}.

%%%%%%%%%%%%%%%%%%%%%%%%%%%%%%%%%%
\section{Background \& Related Work}
\label{sec:related}
%%%%%%%%%%%%%%%%%%%%%%%%%%%%%%%%%

Different terms are used for the privacy decisions involving more than a single individual \cite{humbert2019survey}. We adopt the term \textit{multi-party privacy}~\cite{such2018multiparty} as it captures the cases where the decision on privacy should be taken by multiple individuals involved in a group. The early 2000s saw the rise of the online social networks (OSNs), where the users connect with each other through a defined relationship (e.g., \textit{friendship}) over the network, form groups and share content either publicly or with a limited amount of users within the network. The popularity of the OSNs resulted in a tremendous amount of information shared among their users, thus resulted in the need to define new ways to preserve privacy of the OSN users. Lederer \textit{et al.}~\cite{lederer2004personal} stated that managing roles or groups for privacy in an OSN is a significant burden for the OSN users; therefore, the actions of the users are not always in a privacy preserving manner. Jones and O'Neill~\cite{jones2010feasibility} aim to cluster OSN users to create automated groups and define group-based privacy control without the need of significant amount of user input. While lifting the burden from the users for privacy decisions is a beneficial direction, OSNs are highly dynamic where groups and connections between the users form and dissolve actively, and with the addition of dynamism in contextual properties making the privacy preservation even more difficult to achieve~\cite{olson2005study}.

To overcome the difficulties of preserving privacy in multi-user environments various techniques exist.
Squicciarini \textit{et al.}~\cite{Squicciarini:2009:CPM:1526709.1526780} propose an auction-based model with which OSN users can enter auctions for multi-party privacy management decisions over sharing content. Ulusoy and Yolum~\cite{ulusoy2021collaborative} propose, PANO, which shares the same idea of using Clarke-Tax mechanism in multi-party privacy decisions. These are the most related studies to this work as the game that we propose is based on the same auction mechanism. Yet, \panort differs from them with its additional settings that enable to directly observe the user behaviors within such a mechanism considering both individual preferences and the effect of other users on these preferences.

Akcora \textit{et al.} \cite{akcora2012privacy} use an active learning approach to assign risk levels to sharing content over social media and inform users about how risky it would be to share a piece of content.
Such and Rovatsos~\cite{Such:2016:PPN:2891451.2821512} develop a negotiation-based approach with pre-defined user privacy policies, and the users try to resolve conflicts by making use of their own policies. Such and Criado~\cite{7426841} follow up this work, which contains a software mediator managing the negotiation process without the need of human input, with the help of user modeling algorithms.

K{\"o}kciyan \textit{et al.} \cite{Kokciyan:2017:AAR:3106680.3003434} propose an argumentation based mechanism, with which software agents act on behalf of OSN users to provide arguments for convincing other users to accept their privacy preferences. Misra and Such~\cite{misra2017pacman} develop a personal assistant agent that recommends personalized access control decisions for social media users, based on the social context and utilization of users' social media profiles.
Ilia \textit{et al.}~\cite{ilia2017sampac} develop a collaborative multi-party access control model that makes use of OSN users' social relationship where collective privacy policies can be applied with trusted connections. Such {\it et al.} \cite{such2017photo} investigate the characteristics of shared content which result in multi-party privacy conflicts, from the perspective of how people resolve them, and how the co-owner relationships have affect on these resolutions. Ajmeri \textit{et al.}~\cite{ajmeri2017arnor} use social norms for agent-oriented software engineering methods and propose privacy-aware personal agents that incorporate these norms. Mosca \textit{et al.}~\cite{mosca2020towards} investigate a utility and value-driven approach which considers explainability, role-agnosticism and adaptability in its core.

Colnado \textit{et al.}~\cite{colnago2020informing} conduct semi-structured interviews with participants about using personal privacy assistants in the Internet of Things (IoT) domain. The authors conclude that control over privacy decisions with minimal information overload is the main concern of the privacy assistant users, however, they do not focus on multi-party privacy and do not provide an application as a solution to the aforementioned issues. Abdi \textit{et al.}~\cite{abdi2021privacy} investigate the use of privacy norms with personal privacy assistants in smart home applications. They identify similarities in the contextual integrity parameters for the participants, and conclude that it would be possible to distill general privacy norms from individuals for the use of other privacy assistant users. Feng \textit{et al.}~\cite{feng2021design} study how people exercise their privacy choices in real-life systems, and construct a design space based on a user-centered analysis. The authors also provide a use case in the IoT domain; however, they do not address how privacy choices are affected when the content is co-owned. Windl \textit{et al.}~\cite{windl2023investigating} investigate the use of tangible privacy mechanisms to provide control over smart home sensors and argue that
the privacy mechanisms should provide features such as transparency and awareness with easy-to-manipulate dashboards. However, their study also only focuses on personal privacy and not multi-party privacy. Lee and Weber~\cite{lee2024revealed} conduct a user experiment to investigate the consistency and rationality of privacy choices in social networks, and conclude that even though the people have some privacy awareness, their rationality in sharing decisions is very limited. Zhou \textit{et al.}~\cite{zhou2024bring} introduce an interactive negotiation system with which users can negotiate over privacy configurations of IoT devices. The authors conduct a user study to test the capabilities of the proposed system with $30$ participants, and demonstrate the effectiveness of negotiation-based methods with increased user satisfaction and high agreement rate for resolving privacy conflicts. Despite being focusing on IoT domain, this study can be considered as a complimentary work to our research goals in this paper over online meeting platforms and social networks, where the authors showcase the capabilities of negotiation-based approaches in resolving multi-party privacy conflicts with the results of a user study.

Another domain which recently gained attention for considering multi-user privacy concerns is online meeting platforms. During the Covid-19 pandemic, there has been a surge to have work-related or social meetings using online meeting platforms, which resulted in many privacy violations, such as other people sharing visual or audio content of meeting participants without their consent. Therefore, some researchers started to work on investigating the effects of online meeting platforms on privacy, how to move forward to resolve the issues, and what would be the future directions to take. Kagan \textit{et al.}~\cite{kagan2023zooming} investigate the security and privacy risks in online video conferencing applications. They extract private information from collage images from online meetings over a large dataset, and show that it is relatively easy to collect publicly available images from video conference meetings and extract private information about the participants, such as face images, age, gender and so on. Prange \textit{et al.}~\cite{prange2022saw} conduct an online survey to investigate the experiences of online meeting users where they faced privacy-invasive situations. The results show that online meetings caused private information to be revealed more than physical meetings, and not just of the participants, but also of their children or spouses. Cardon \textit{et al.}~\cite{cardon2023recorded} study a different aspect of recorded meetings, where AI tools can be used to analyze and evaluate the meetings for the benefit of the businesses. The results show that even though AI tools can be useful from the business perspective, they also cause issues about privacy and safety violations and give unwarranted control to employers over employees. Sandhu \textit{et al.}~\cite{sandhu2023unfolding} provide a conceptual research model based on privacy calculus and the privacy concerns of the video conferencing application users. The results indicate that privacy protection should be aimed at organizational level in businesses and control mechanisms should be in place to motivate employees to actively engage in privacy protection behavior. These studies strengthen the idea to have privacy preserving mechanisms for collaborative systems, which we also aim to investigate in this paper.

\section{User Study}\label{sec:user_study}

The aim of this research is to simulate and understand the preferences of users when sharing an image that depicts several people. We want to understand how other people's preferences impact individual desire to share or not share an image of themselves and if people take choices that contract group's preference. In particular, we want to answer the research questions given in Section~\ref{sec:intro}.

\subsection{RESOLVE: A game for multi-party privacy}
 In order to investigate these research questions, we first devised a game~\panort with a group-decision mechanism. The game enables players who appear in a given group picture express how much they are willing to share or not share the picture individually and then combines individual choices to decide whether the group picture will be shared or not. After the first group-decision is obtained by the combination of these choices, players are asked to reevaluate their choices, considering the preferences of others and the current outcome.

\subsubsection{Choice of the group-decision mechanism}
In order to reach the group-decisions, we considered several mechanisms that are based on argumentation, negotiation, as well as auctions. One key requirement was to have a mechanism that is easy to understand by the participants, without requiring too much technical knowledge. Argumentation-based approaches~\cite{Kokciyan:2017:AAR:3106680.3003434}, while powerful, require participants to process too many details, including logic-based reasoning, to take part in the mechanism.
The negotiation-based mechanisms are also powerful as they enable participants to express what aspects of privacy are more important for them and enable them to make offers and counter-offers accordingly. The negotiation-based mechanisms are interactive and might require participants to continue interactions for a number of iterations. While in some situations this is desirable, in the context of our research questions, it is difficult to foresee the effect of these iterations. The auction-based mechanisms, on the other hand, are simple as they do not require participants to interact with each other. Each participant votes for the share or not-share decision and expresses a valuation in terms of a bid that demonstrates how much they value the decision. Based on the above constraints, we choose to use PANO \cite{ulusoy2021collaborative}, an auction-based group-decision mechanism and gamify it with the details below to realize \panort.

\subsubsection{Workings of RESOLVE}\label{sec:auction_mechanisms_in_multiparty_privacy}
Each player is shown a group picture and is asked to state their privacy choice on the picture.

\mypara{Choice of outcome:} When a player is shown a group picture, they are asked to choose between \textit{``Share''} or \textit{``Not Share''} outcomes and to make a {\it bid}, which demonstrates how strongly they would like the outcome (i.e., the higher the bid, the more they would like to see the outcome). PANO enables its users to easily build bidding strategies by limiting the bid amount for an auction. Similarly, we allow the player to make a bid between 1 and 20 in each game. Complying with the feature of PANO for group-wise spending, every player has a {\it budget} that reflects the total amount of credits that they can use for the overall game. Each player starts with a budget of $50$ and obtains an extra $10$ credits for each content that they need to bid. The goal of giving $10$ for each game, which is the half of the maximum bidding boundary, is to make the participant think carefully about their choices and bid with care. Always bidding high would make them not saving up, which would eventually lead to the participant not having enough budget to bid closer to the high boundary. In the meantime, bidding a low amount might cause them to not get a favorable outcome in the current auction, but they would save budget to have a say in a future game. In this manner, concerns about fairness in overall decisions that people might have in a social setting are also addressed in the game.

\mypara{Rethinking the choice:} After each user stated their preferences and their bids to support them (first round), the total bids for each choice are calculated. The choice with the greater total amount of bids becomes the overall decision of the group. Before announcing the definite result, the game displays the preferences of others (without their bids) to give the opportunity to the player to rethink their choice (second round). At this stage, the participant has an option to change their choice (Share or Not Share) as well as their bid. This enables us to observe various user behaviors, including fighting for their preferences and accommodating others.

\mypara{Outcome of the game:} The bids for each choice from all players are summed once again. The players who preferred that choice become winners of the game and the resulting decision is called a \textit{favorable outcome} from their point of view. In order to discourage the same individuals to always influence the outcome, a winning player of the auction is \textit{taxed} a fixed amount of $5$ if the overall decision would be different without the player. In this way, the rational action for each player becomes to bid on their actual choice, and they cannot always be in an advantageous position for the game with this extra deduction from their budget~\cite{RePEc:kap:pubcho:v:11:y:1971:i:1:p:17-33}. If a player wins the game, the player's {\it payment} is the sum of the amount that they have bid for the content and the tax, if they have been taxed. If the player loses the game, their payment will be zero.

\mypara{Rewarding of participants:}
 A participant was {\it rewarded} based on two points: favorable outcomes in the games and saving up budget throughout the experiment. To determine the favorable outcomes that we can reward, we used the games where the participants won both rounds with their choices (Share or Not share). Saving up the budget implies that the participant i) placed their bids carefully to not overspend, and ii) did not bid high values, therefore, forwent their individual choices for the sake of the choices of the others. This would also result in her not paying taxes when the bid of the participant do not change the outcome of the game, meaning that the participant gave up their individual choice for the benefit of the society. With these two items (favorable outcomes, saved budget), we define rewards as 0.50 for each favorable outcome and 0.10 for each $10$ credits of the budget saved in the end so that winning the game is valued over the saved budget.

\subsection{Procedure}\label{sec:proceduere}

At the start of the user study, the participant was directed to a website where they would be explained about the steps of the study and read the consent. Appropriate approvals were obtained from the Ethics Board of the institution administering this study and this information was also given to the participants.  In order to ensure that they understand the requirements of the study and their rights, they are asked to check multiple statements about data usage. Once the consent was signed, the participant was directed to register and create an account. The account enabled tracking of the progress of each participant and allowed each participant to pause the experiment and resume later. While creating an account by choosing a username and password, the participant was also asked to provide their demographic information (age, gender, education level, income and ethnicity), sharing behaviors and concerns about privacy. The responses of the latter two questions were collected in a 5-point Likert scale.

\begin{figure}[H]
    \centering
    \includegraphics[scale=0.73]{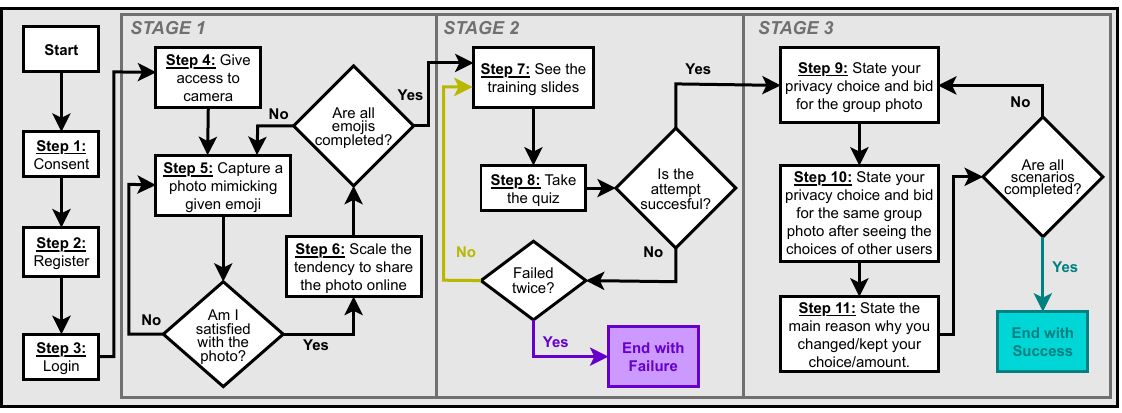}
    \caption{The flow of the user study}
    \label{fig:experiment_flow}
    \Description{Flowchart consisting of start state, eleven steps of the study, and two end states, one with failure and one with success. Steps are included in corresponding stages for which details are provided in the paper itself.}
\end{figure}

Following this preliminary phase, the study consists of three main stages (see Fig. \ref{fig:experiment_flow}):
\begin{enumerate}
    \item {\bf{Capturing Pictures}}: During this stage, the participant was asked to take pictures mimicking an emoji shown on the screen and to state their tendency to share for each image.
    \item {\bf{Training}}: During this stage, the participant was familiarized with the different aspects of the experiment. This phase was used to train the participant and to test their understanding with a quiz. Participants who failed in the quiz were asked to retake the training. After failing the quiz twice they could not longer take the experiment.
    \item {\bf{Experiment}}: This phase is the experiment phase where we measured the preferences and motivations of the participant when sharing their information online through the game we designed.
\end{enumerate}

\subsubsection{Capturing Pictures}

\begin{figure}[H]
    \centering
    \includegraphics[scale=0.23]{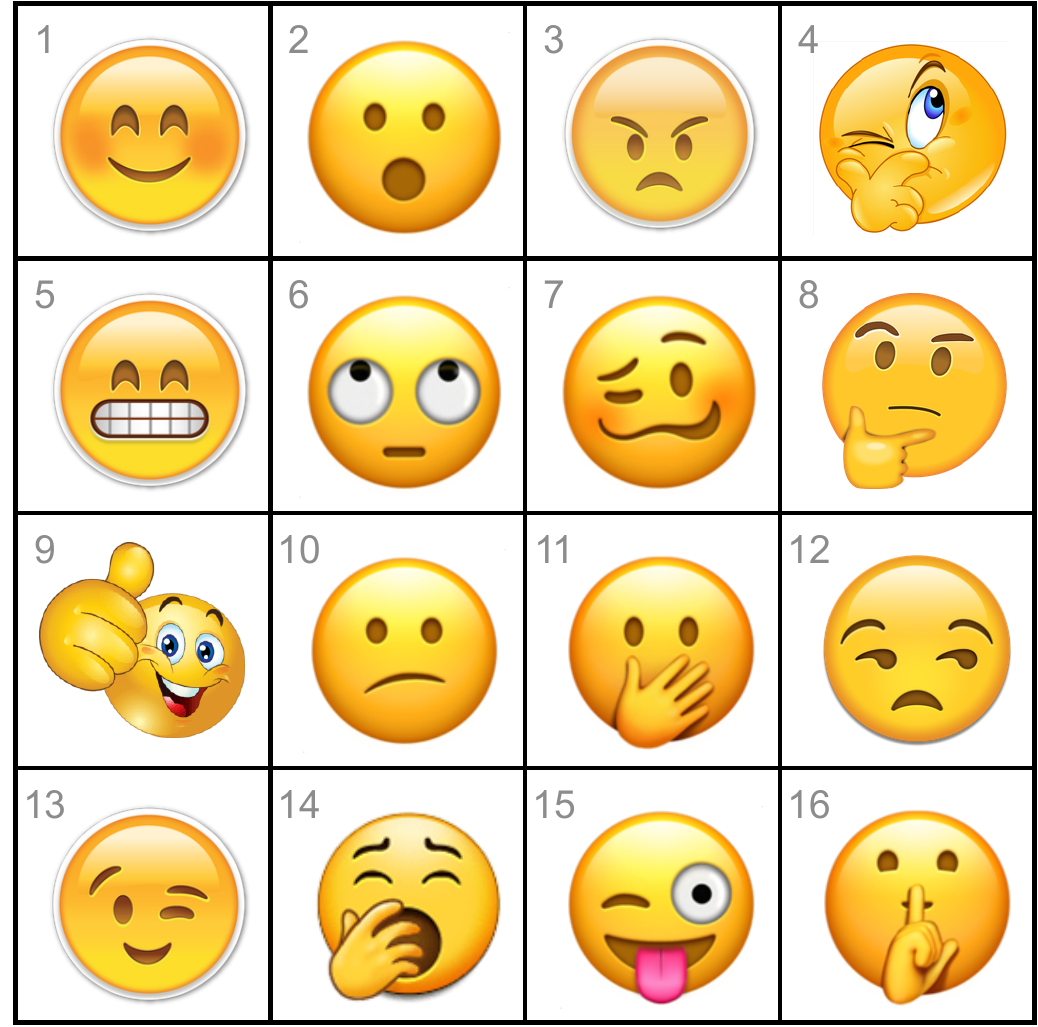}
    \caption{The set of emojis used in the experiment with their numerical labels.}
    \label{fig:emoji_set}
    \Description{The set of emojis used in the experiment with their numerical labels.
    1) Smiling Face with Smiling Eyes
    2) Face with Open Mouth
    3) Angry Face
    4) Not Believing Face
    5) Grimacing Face
    6) Face with Rolling Eyes
    7) Woozy Face
    8) Thinking Face
    9) Face with Thumb Up
    10) Slightly Frowning Face
    11) Face with Hand Over Mouth
    12) Unamused Face
    13) Winking Face
    14) Yawning Face
    15) Winking Face with Tongue
    16) Shushing Face}
\end{figure}

\begin{figure}[h!]
    \centering
    \subfloat[]{
        \includegraphics[scale=0.27]{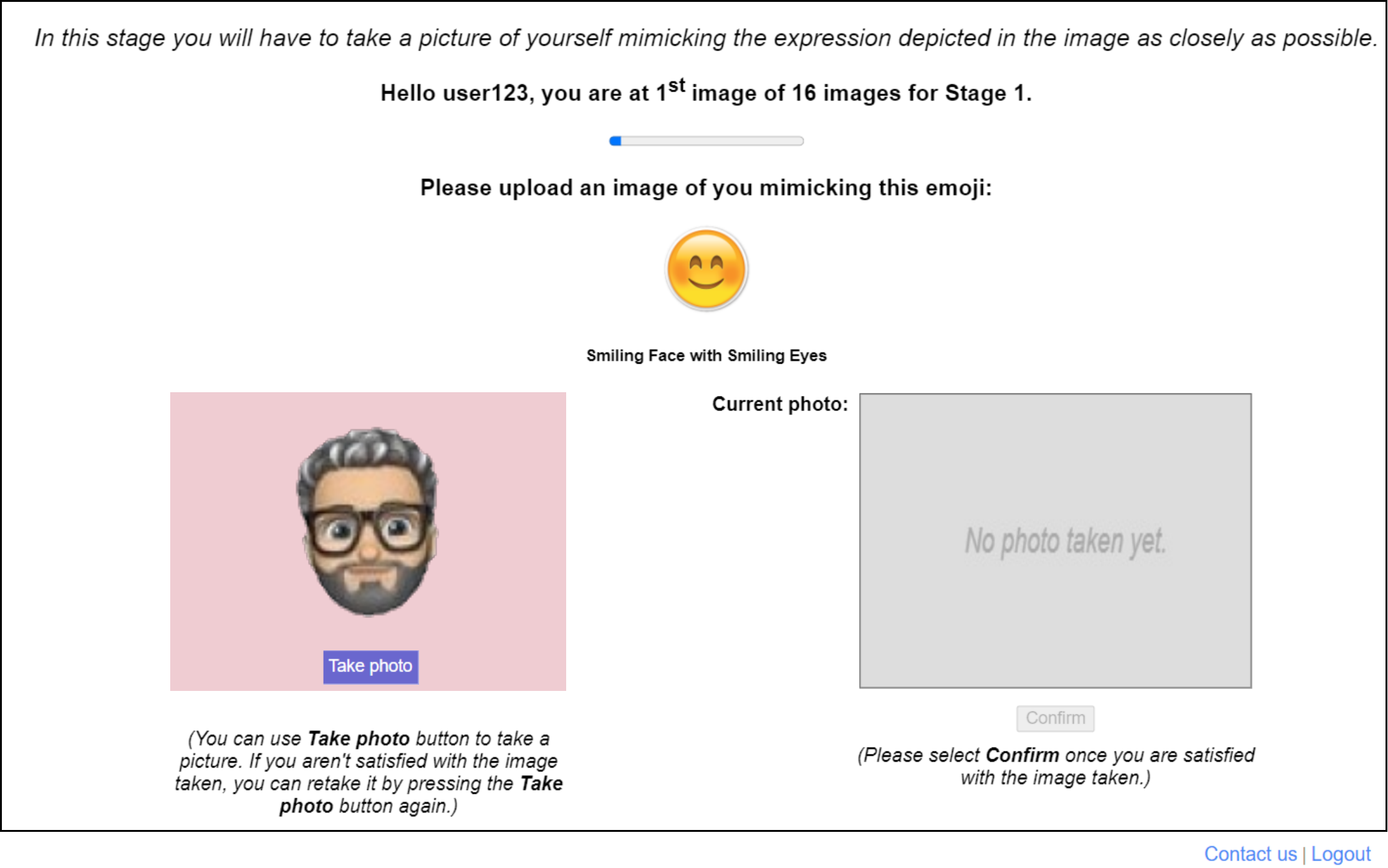}
        \label{fig:webcam_sample}
    } \\
    \subfloat[]{
        \includegraphics[scale=0.27]{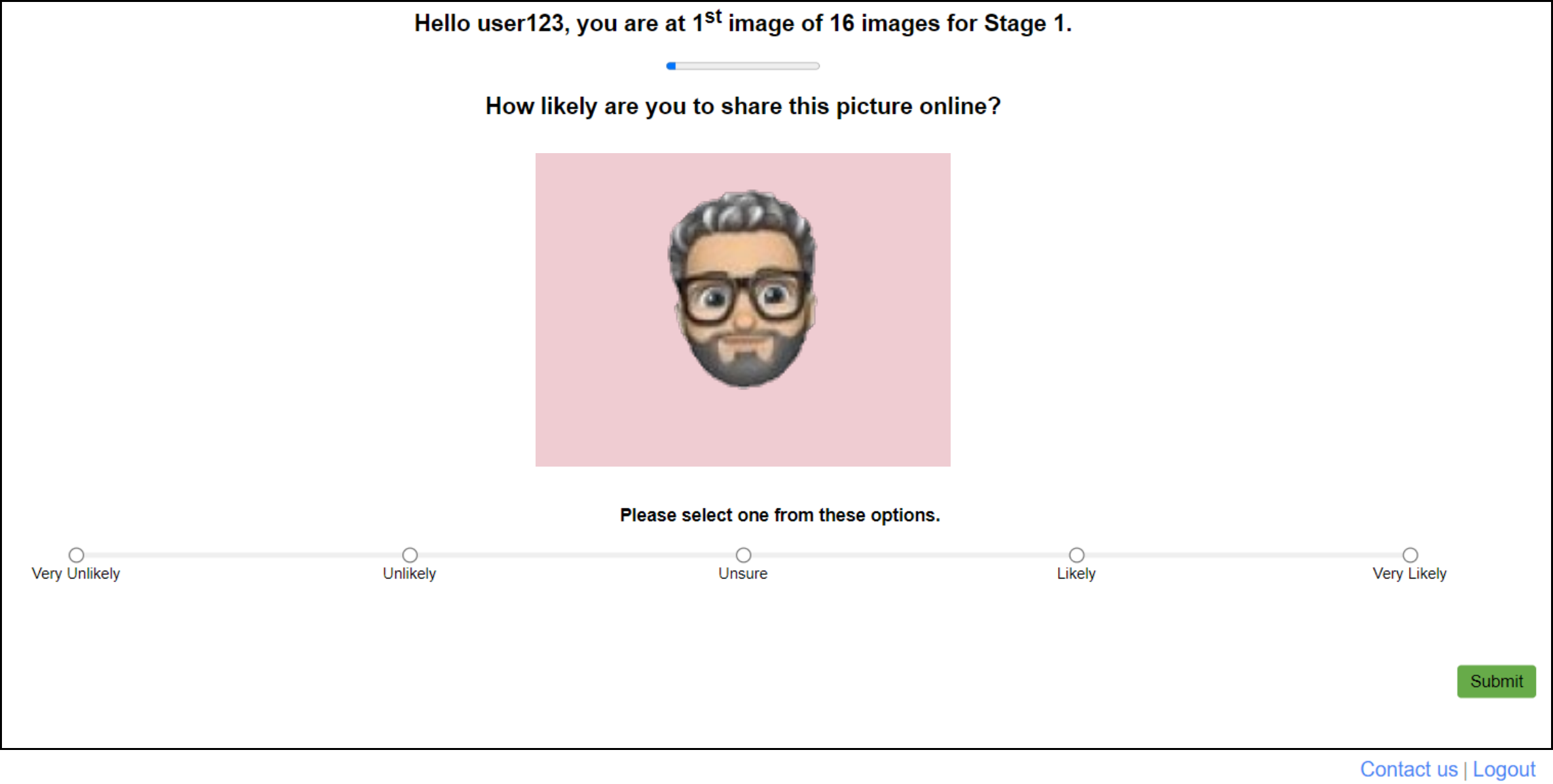}
        \label{fig:likert_sample}
    }
    \caption{Sample images for Step 5 (a) and Step 6 (b). Here, the picture of the participant is replaced with a memoji to illustrate the interface anonymously. }
    \label{fig:sample_step5_and_step6}
    \Description{Sample image for the interfaces of (a) Step 5 - Capturing own image while mimicking the given emoji, and (b) Step 6 - Stating tendency to share the photo taken in Step 5. Upper subfigure (a) includes two rectangle areas, one for reflecting live image of the camera which has also a button labeled "Take photo", and the other one for the last photo taken by the user. Below to the latter area, there is confirmation button for the user to upload the current photo, if the user is satisfied with it. The subfigure below (b) has the area showing the photo uploaded by the user, and a horizontal radio button group to select the tendency to share in 5-point Likert scale.}
\end{figure}

In this stage, the participant was first asked to enable their camera.  Then, the participant was shown an emoji and was asked to take a photo when mimicking the emoji. The participant was free to take as many shots as they like before uploading the one they prefer. After receiving a confirmation of successful upload, the participant was asked to rate the photo considering how likely they would like to share it online using a Likert scale, with the following five strengths: namely \textit{Very Unlikely}, \textit{Unlikely}, \textit{Unsure}, \textit{Likely}, and \textit{Very Likely}. Figure \ref{fig:webcam_sample} and Figure \ref{fig:likert_sample} show an example participant mimicking an emoji and ranking the photo respectively. The participant was asked to repeat these for $16$ different emojis (see Figure \ref{fig:emoji_set}).  The same set of emojis were shown to all participants. A progress bar was shown at the top of the screen to indicate the participant’s progress in the study.

\begin{figure}[H]
    \centering
    \includegraphics[scale=0.44, frame]{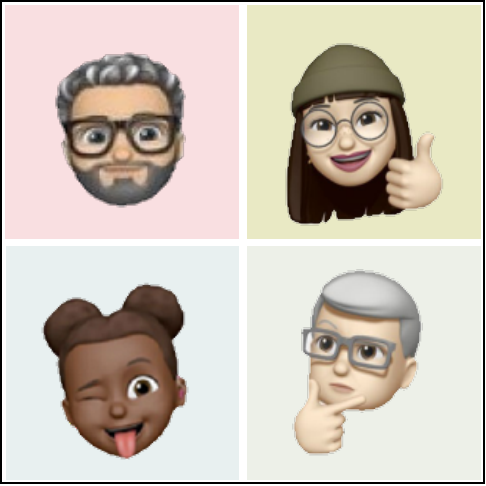}
    \caption{Sample illustration with memojis for group pictures created in the experiment.}
    \label{fig:sample_zoom}
    \Description{Sample group picture created in the experiment resembling a screenshot from an online meeting where the screen is divided into four, each area showing one individual.}
\end{figure}

We created $16$ group pictures such that for each group picture one photo of the participant was concatenated to a prepared picture that shows three other individuals.
These group pictures resemble an online meeting, such as Zoom, where the photos are structured in a grid format. This creation happens in the background, and the participant was not shown the group picture at this stage. At the end of the participant's session, we had $16$ group pictures where the participant was also present. Fig.~\ref{fig:sample_zoom} shows an example of such a group picture.  These group pictures were used not only in the experiment but also to introduce the game more realistically for the training of the participants in the next stage.

\subsubsection{Training}
Stage 2 starts with training of the participant where a set of slides are shown to them. These slides provides the necessary information for the experiment to the participant. Hence, the participant learns about the interface and the game including:

\begin{itemize}
    \item how the scenarios that require a common decision are constructed and how they can state their own privacy choice and bid to support it for each of those scenarios,
    \item how the choices of others are shown for the second round of each scenario,
    \item how they can see the progress they have made so far in terms of numbers of participated scenarios and favorable outcomes in those scenarios, and their current reward,
    \item how the payment for a scenario is calculated based on the choices and bids of the whole group,
    \item how their overall performance in the experiment is evaluated based on the budget and the number of favorable outcomes.
\end{itemize}

The participant was able to freely navigate among the slides, but the button to take the quiz became available once all slides have been seen. A button which makes the same set slides available in a separate tab was provided at the bottom of the interface in case she needs to revisit them during the quiz and the execution of the game. After the quiz was initiated by the participant, the participant was required to answer ten questions in different categories. Further information for these categories is provided in Section \ref{sec:validity_and_limitations}. The complete text for the quiz questions along with the training materials are provided in Appendix \ref{sec:training_material}.

If the participant failed to pass the quiz, they were redirected to training slides where they got a second chance to have a look at the slides and retake the quiz. In case of a second failure, they were redirected a page stating they were not eligible for the experiment and they need to send an email to the researchers if they still want to retake the quiz and participate in the experiment.

\subsubsection{Experiment}
\label{sec-experiment}

If the participant passes the quiz, they start playing the game of \panort with three software agents, which are controlled by our system. The scenarios that were constructed with the group images as shown in Figure \ref{fig:sample_zoom} are shown to them one by one, in a randomized order.

\begin{figure}[h!]
    \centering
    \subfloat[]{
        \includegraphics[scale=0.44]{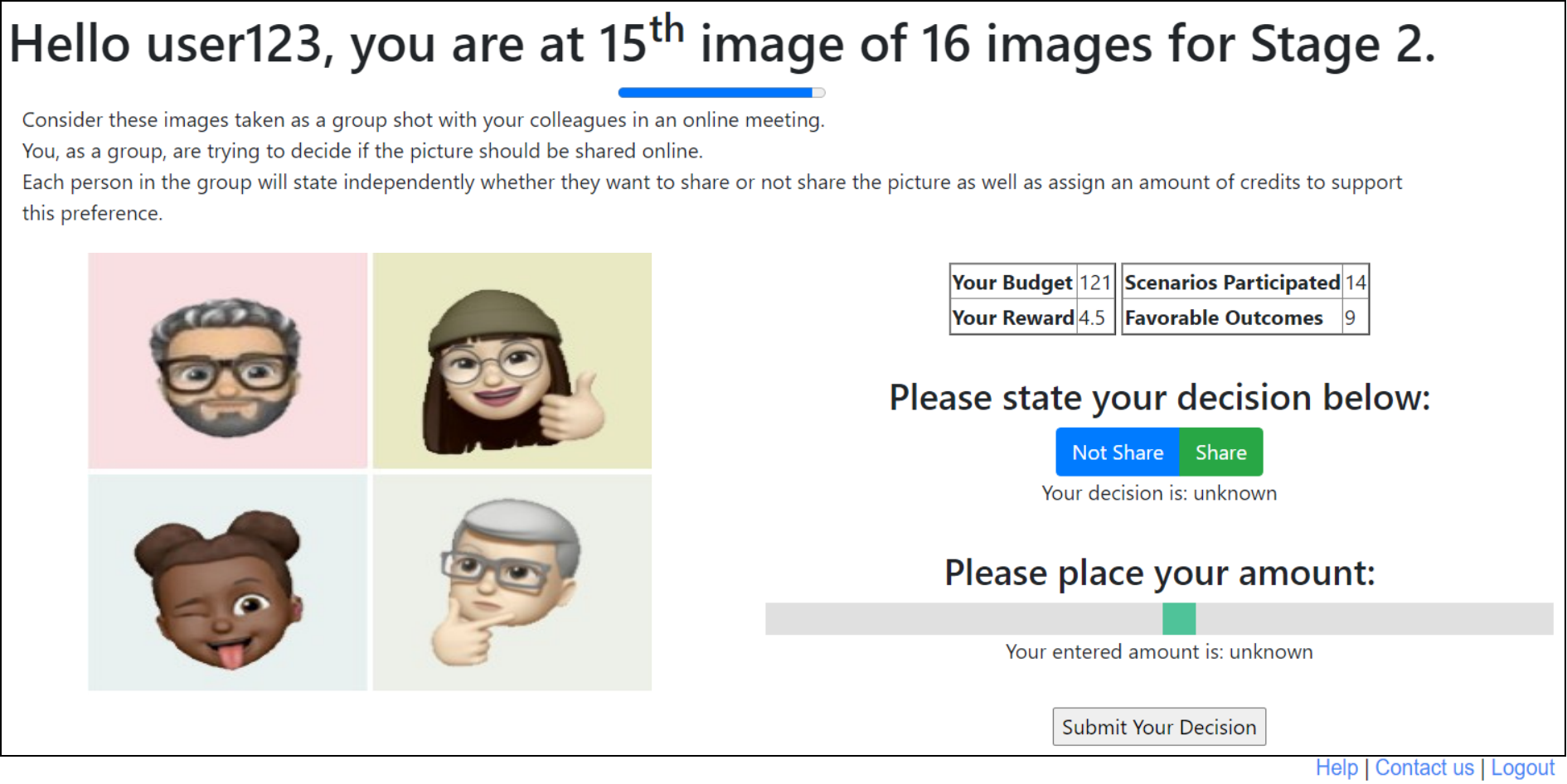}
        \label{fig:sample_1st_round}
    } \\
    \subfloat[]{
        \includegraphics[scale=0.38]{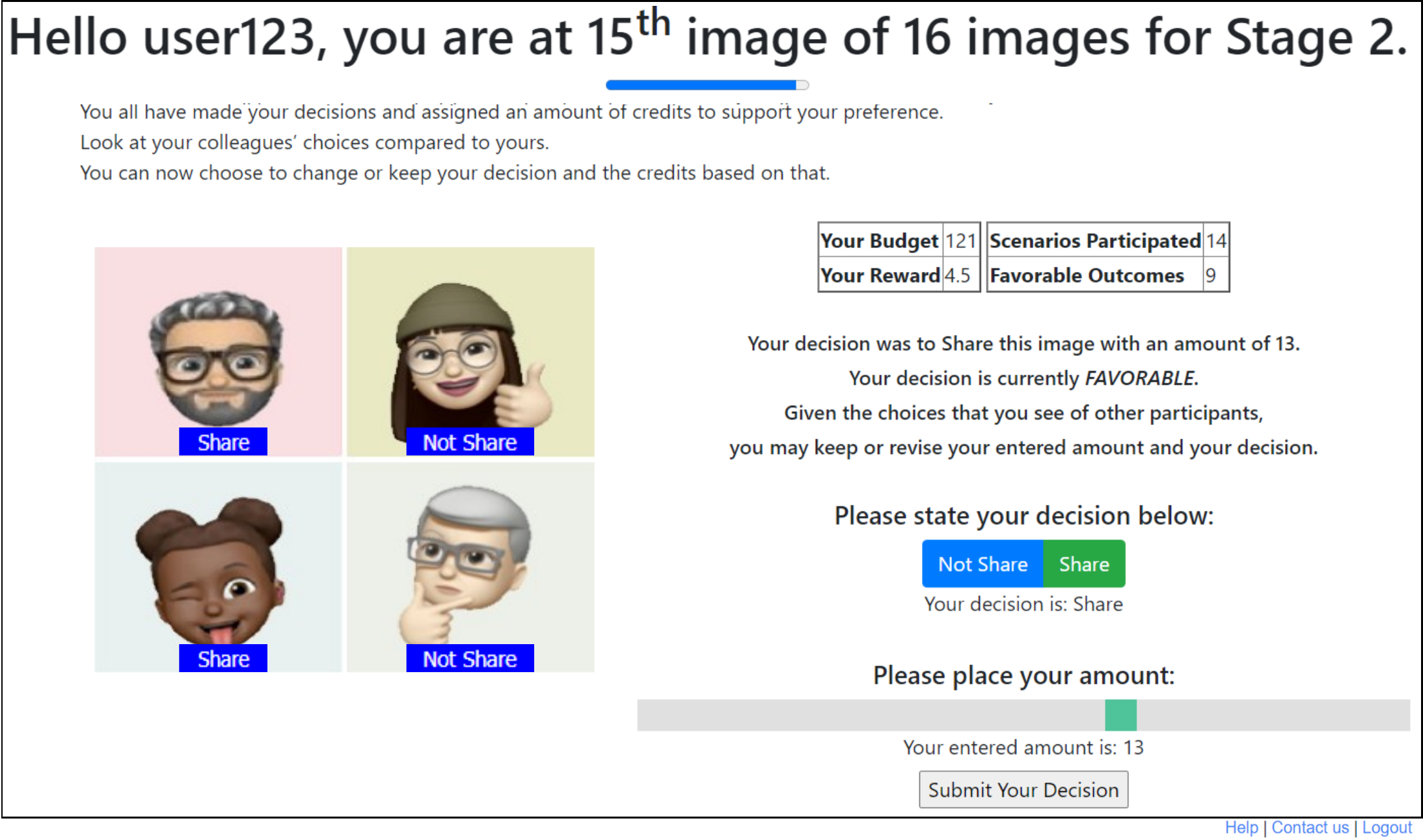}
        \label{fig:sample_2nd_round}
    }
    \caption{Sample illustration for the game in two rounds: Before (a) and after (b) seeing the choices of others.}
    \label{fig:sample_rounds}
    \Description{Sample screenshots from the interface for the game in two rounds. For the first round, left part of the screen is used to show the group picture used in the current scenario. Right part shows a table for the current performance of the participant, two button to select "Not Share" or "Share" as their privacy choice, and a slider to set their bids for that choice. Below to all, there is a button to submit individual decision. Almost same interface is arranged for the participant to indicate their choice and bid in second round, however, the previous privacy choices of all individuals are embedded in the group picture. Also, the outcome as the group-decision on the picture of previous round is provided in the right part this time.}
\end{figure}

Each scenario required the participant to state their privacy choice and bid twice, before and after they were able to see the choices of others and the current outcome (see Fig. \ref{fig:sample_rounds}). Another progress bar was available on top of the study to show the participant’s progress in this stage as well.

As the last step of each scenario, the participant was asked for the main reason why they kept their individual decision as their choice and bid, or changed at least one of them. Four predetermined expressions for each case were provided them along with an ``Other'' option to enable them to write their motivation in their own words (see Fig. \ref{fig:all_behaviors_p2} for the predetermined expressions).

Once all of the scenarios have been completed, the participant was redirected to the final page where they could see the number of scenarios where they changed their choice or bid. Their final budget, the number of favorable outcomes in the experiment, and how many tickets for the lottery as a reward they have got based on those two were also reported in this page.

\subsection{Scenario Designs}\label{sec:scenario_designs}

We had $16$ scenario permutations that aim to capture various real-life cases to measure how participants would act in different situations in relation with multi-party privacy decisions. Each scenario had an image of the participant, along with images of three other humans. However, the bidding for these images were done by software agents in the background in a predetermined manner as explained below.

\clearpage

Scenarios were created with the images that participant captured by themselves imitating $16$ different emojis. They also stated their likeliness to share for each image in Likert scale, from \textit{``Very Unlikely''} to \textit{``Very Likely''}. Based on this input, we categorized the images as \textit{``Unlikely''}, \textit{``Unsure''} and \textit{``Likely''} by considering extreme ends as the same with the corresponding intermediate options. As our aim is to cover all possible combinations for the individual choices of both the participant and the software agents, we used two image categories, \textit{``Unlikely''} and \textit{``Likely''} which presumably lead to the choices \textit{``Not Share''} and \textit{``Share''}. Table \ref{tab:scenariosLikert} shows the ideal case for the categories of images that would be used in the scenarios.

In an ideal setting, each participant would have eight images for each type of sharing tendencies as shown in Table \ref{tab:scenariosLikert}. Yet, as a more realistic approach,  we had to consider the case where a participant does not provide enough number images that belong to a category that fit into the corresponding scenarios. To compensate such a case we applied the following selection criteria for the images in the given order:

\begin{itemize}
    \item If the participant has another image in the proper category for the scenario, then the same image is used again.
    \item If the participant has an image that they are \textit{``Unsure''} about sharing, then the image is used in the scenario.
    \item If the participant has no image in the same category or the category of \textit{``Unsure''}, then a random image is chosen from the opposite category for the scenario.
\end{itemize}

\begin{table}[h!]
    \scalebox{0.95}{
    \begin{tabular}{|c|c|c|c|c|}
        \hline
        \multirow{2}{*}{\textbf{Scenario}} & 
        \multicolumn{4}{c|}{\textbf{Image Category for Each Individual}} \\
        \cline{2-5}
        & \textbf{Participant} & \textbf{Agent 1} & \textbf{Agent 2} & \textbf{Agent 3}\\
        \hline
        1   & Unlikely    &Likely&   Likely&   Likely\\
        2   & Unlikely    &Likely&   Unlikely&   Likely\\
        3   & Unlikely    &Unlikely&   Likely&   Unlikely\\
        4   & Unlikely    &Unlikely&   Unlikely&   Unlikely\\
        5   & Unlikely    &Likely&   Likely&   Likely\\
        6   & Unlikely    &Likely&   Unlikely&   Likely\\
        7   & Unlikely    &Unlikely&   Likely&   Unlikely\\
        8   & Unlikely    &Unlikely&   Unlikely&   Unlikely\\
        9   & Likely    &Likely&   Likely&   Likely\\
        10   & Likely    &Likely&   Unlikely&   Likely\\
        11   & Likely    &Unlikely&   Likely&   Unlikely\\
        12   & Likely    &Unlikely&   Unlikely&   Unlikely\\
        13   & Likely    &Likely&   Likely&   Likely\\
        14   & Likely    &Likely&   Unlikely&   Likely\\
        15   & Likely    &Unlikely&   Likely&   Unlikely\\
        16   & Likely    &Unlikely&   Unlikely&   Unlikely\\
         \hline
    \end{tabular}}   
    \caption{The images of the scenarios in the ideal case, according to categorization based on the Likert scale input about how likely would the participants share the content.}
    \label{tab:scenariosLikert}
    \Description{Categories for Individual Images of Each Player. As possible categories are "Unlikely" and "Likely", we have 16 different scenarios which covers all possible combinations.}
\end{table}

Nevertheless, we assume that in half of the scenarios participant would like to share the content, while they would prefer keeping the content private in the remaining half leading to expected choices given in Table \ref{tab:scenariosDecision}. One can notice that the expected behavior of the participant in each scenario is coherent with the tendencies given Table \ref{tab:scenariosLikert}, yet the choices of the software agents vary in Scenario 5 to 8 and in Scenario 13 to 16. This contradiction is applied as a reflection of real-life behavior where people may want to share an image, they preferred keeping it private earlier. On the other hand, people may also want to keep an image private, while they would like to share initially, either due to lack of knowledge or considering other factors such as the opinion of the society.

Overall, through the diverse behaviors of three software agents, we generate 16 different scenarios that cover both the agreement and disagreement of the participant with the majority of the group. These scenarios can be categorized as follows:

\begin{itemize}
    \item The participant does not want to share the group photo, while the majority wants it to be shared (\textit{Scenario 1, 2, 7, and 8}).

    \item The participant does not want to share the group photo, neither do the majority of the people in it (\textit{Scenario 3 to 6}).

    \item The participant wants to share the group photo, while the majority wants it to be kept private (\textit{Scenario 11 to 14}).

    \item The participant wants to share the group photo, as well as the majority of the people in it (\textit{Scenario 9, 10, 15, and 16}).
\end{itemize}

\begin{table}[h!]
\scalebox{0.95}{
    \begin{tabular}{ |c|c|c|c|c|}
        \hline
        \multirow{3}{*}{\textbf{Scenario}} & 
        \multicolumn{4}{c|}{\textbf{Decisions for Each Individual}} \\
        \cline{2-5}
        & \textbf{Participant} & \multirow{2}{*}{\textbf{Agent 1}} & \multirow{2}{*}{\textbf{Agent 2}} & \multirow{2}{*}{\textbf{Agent 3}}\\
        & \textbf{(Expected)} & & & \\
        \hline
        1   & Not Share    &Share (5) &   Share (5) &   Share (5)\\
        2   & Not Share    &Share (10) &   Not Share (10) &   Share (10) \\
        3   & Not Share    &Not Share (5) &   Share (5) &   Not Share (5) \\
        4   & Not Share    &Not Share (5) &   Not Share (10) &   Not Share (10) \\
        5   & Not Share    &Not Share (10) &   Not Share (10) &   Not Share (15) \\
        6   & Not Share    &Not Share (10) &   Share (15) &   Not Share (5) \\
        7   & Not Share    &Share (5) &   Not Share (5) &   Share (10) \\
        8   & Not Share    &Share (5) &   Share (5) &   Share (10) \\
        9   & Share    &Share (5) &   Share (5) &   Share (5) \\
        10   & Share    &Share (5) &   Not Share (10) &   Share (5) \\
        11   & Share    &Not Share (10) &   Share (5) &   Not Share (5) \\
        12   & Share    &Not Share (10) &   Not Share (10) &   Not Share (10) \\
        13   & Share    &Not Share (5) &   Not Share (5) &   Not Share (5) \\
        14   & Share    &Not Share (15) &   Share (10) &  Not Share (5) \\
        15   & Share    &Share (5) &   Not Share (15)&   Share (5) \\
        16   & Share    &Share (10) &   Share (15) &   Share (15) \\
         \hline
         
    \end{tabular}}
    \caption{The decisions of the software agents in the scenarios with the amount of bids given in parentheses and the expected decision of the participant according to their input about how likely they would share the image.}
    \label{tab:scenariosDecision}
    \Description{Expected decision of participant and fixed decision and bids for the agents in the game. In order to create all possible outcomes with each privacy choice, decisions are mostly fixed complying to image categories, yet some random decisions are applied in the agents to cover the randomness in real users. Scenario types based on these decision combinations are covered in the paper itself.}
\end{table}

\subsection{Apparatus}\label{sec:apparatus}

\panort serves as the main apparatus, where the interface was developed as a Web Application by using \texttt{Java Servlet} framework while \texttt{Google Chrome} was selected as the main browser for testing. Accordingly, the participants were asked to access the study through this browser to avoid any compatibility issues.

A logout functionality was provided to the participant so that they could pause in anytime and login again using the username and password created during the registration. These credentials and other demographic information of the participant were kept in a relational database along with the data produced by the participant through the study. The username was used during the study to address the participant, so that they can consider the interface as the interface of a real OSN. This username was also used as one of the main primary keys in the database to associate different aspects of the participant's data. However, a separate user id was assigned to each participant so that the username can be replaced with it during the preprocessing of the data for the sake of participant's anonymity.

The photos uploaded by the users were kept in a local folder of the server until the analysis was done. These images were manually validated so that we could ensure that participants actually imitated the given emojis and took their parts in the scenarios as they were asked to. Nonetheless, other than this validation process, the images were not included in the analysis and deleted afterwards.

Despite being excluded in the analysis of experiment results, these group photos were crucial in the scenarios as they enable the participants consider the given scenarios realistic. In order to facilitate this, we created a single content by merging the photo of the participant with the photos of the software agents to imitate a screenshot from an online meeting for each scenario.

\subsection{Recruitment}\label{sec:recruitment}

Participants were recruited using internal mail lists and personal networks. Participants were offered €5 Amazon voucher if they successfully completed the study. Total reward a participant obtained by completing the study was converted to lottery tickets via multiplying it by 10 where participant got the chance to win an extra voucher of €20. Five participants earned this additional voucher. The details about the performances of the participants and the lottery can be found in Appendix \ref{sec:participant_performances}.

\subsection{Participants}\label{sec:participants}

For the study, we recruited $46$  online participants with the mean age of 33.4 ($\pm 9.2$); 22 females with the mean age of 30.9 ($\pm7.7$), 24 males with the mean age of 35.7 ($\pm9.9$). Majority, 39 have a full or part-time job, where the rest is either student, looking for a job or self-employed. Participants are highly educated as 32 of them have a master's or doctoral degree, 10 of them have associate or bachelor's degree and 4 participants are undergraduate students. 27 participants responded as they are \textit{``concerned''} or \textit{``very concerned''} to the question of \textit{``How concerned are you about your privacy on online social networks?''}, where 14 participants have a neutral opinion for the matter, and the rest are either \textit{``unconcerned''} or \textit{``very unconcerned''}. Although the majority of the participants tend to be concerned for their privacy, only one participant responded as \textit{``never''} to the question of \textit{``How often do you post on online social networks?''}. Majority as 25 participants declared their sharing frequency as ``rarely''. ``Occasionally'' and ``frequently'' was chosen by 13 and 7 participants respectively, while there is no participant who stated that they share on OSNs ``very frequently''. Table \ref{tab:participant_demography} in Appendix \ref{sec:participant_characteristics} provides the complete distributions for the characteristics of the participants presented above.

$41$ out of $46$ participants passed the quiz. One of these participants is excluded from the study as their responses could not be validated with the attention check we applied at the end of the experiment (see Section \ref{sec:validity_and_limitations} for details). The remaining $40$ participants successfully completed the experiment by taking part in all 16 scenarios. We use their data in the rest of the paper to answer our research questions.

\section{Content Sharing Behavior in Multi-Party Privacy Decision Making}\label{sec:profiles_in_multiparty_privacy}

Our first research question (RQ1) is to understand different content sharing behaviors of users in the context of multi-party situations. There are three specific sub-questions we are interested to answer. First sub-question is related to users' initial willingness to share their own pictures. To what extent would users share their pictures individually? (Section~\ref{sec-willingness}). Although this question is not directly related to the group context, learning the preferences of users for different individual contents is important to analyze their behaviors regarding the group-decision. The second sub-question is related to users' discrepancy in their expression of privacy and their attitudes. To what extent do users not share the content that they identify as private and vice versa? (Section~\ref{sec-discrepancy}). The last sub-question is related to a typology of users. Can we create user profiles based on their sharing behavior? (Section~\ref{sec-profiles}).

\subsection{Willingness to Share}
\label{sec-willingness}

In order to understand to what extent users are willing to share their content, we study their average tendency to share. Table \ref{tab:participant_avg_tendencies_for_emojis} shows the average likeliness of each participant to share the contents where they mimic the given set of emojis in Likert scale ($L_1$ \textit{- ``Very Unlikely''} to $L_5$\textit{ - ``Very Likely''}). The majority of the participants indicated that they would not be too willing to share the content by assigning low scores to declare their sharing tendencies, leading to low averages with a small variance.  The overall average of given scores is 2.0, where the standard deviation is 0.6. Yet, we also observe for a few participants, a higher average (i.e., close to 3 with higher variance), indicating that they expressed they were more likely to share the images.
Table \ref{tab:participant_tendencies_for_emojis_full_table} in Appendix \ref{sec:tendencies_all_emojis} provides the complete set of tendencies of each participant for sharing the content with all emojis.  These tendency values for each of the 40 participants are used in generating the $16$ scenarios as explained in Section~\ref{sec:scenario_designs}.  Overall, we have 640 cases (scenario instances) in total, where the distribution of the tendencies is skewed to the lower side of the average (see the last row in Table \ref{tab:tendencies_and_decision_pairs}). This also suggests that the participants prefer more privacy for their contents.

\begin{table}[h]
    \centering
    \scalebox{0.78}{
    \begin{tabular}{V{4}c|cV{4}c|cV{4}}
        \hlineB{4}
        \multirow{2}{*}{\textbf{Participant}} & \textbf{Average Tendency} & \multirow{2}{*}{\textbf{Participant}} & \textbf{Average Tendency} \\
        & \textbf{to Share} & & \textbf{to Share} \\ \hlineB{4}

         \cellcolor{myredfor2}
P1 & \cellcolor{myredfor2} 2.7 (0.9) & \cellcolor{myredfor1}
P23 & \cellcolor{myredfor1} 1.5 (0.7) \\ \hline

\cellcolor{myredfor1}
P2 & \cellcolor{myredfor1} 1.1 (0.5)& \cellcolor{mycyanfor3}
P24 & \cellcolor{mycyanfor3} 3.4 (1.1)\\ \hline

\cellcolor{myredfor1}
P3 & \cellcolor{myredfor1} 1.8 (1.2) & \cellcolor{myredfor1}
P25 & \cellcolor{myredfor1} 1.9 (0.7)\\ \hline

\cellcolor{myredfor2}
P4 & \cellcolor{myredfor2} 2.6 (1.8) & 
\cellcolor{myredfor1}
P26 & \cellcolor{myredfor1} 1.1 (0.3)\\ \hline

\cellcolor{myredfor2}
P5 & \cellcolor{myredfor2} 2.0 (1.3) & \cellcolor{myredfor2}
P27 & \cellcolor{myredfor2} 2.4 (1.4)\\ \hline

\cellcolor{myredfor2}
P6 & \cellcolor{myredfor2} 2.8 (1.1) & \cellcolor{myredfor2}
P28 & \cellcolor{myredfor2} 2.0 (1.3)\\ \hline

\cellcolor{myredfor2}
P8 & \cellcolor{myredfor2} 2.0 (0.6) & \cellcolor{myredfor2}
P29 & \cellcolor{myredfor2} 2.6 (1.3) \\ \hline

\cellcolor{myredfor1}
P10 & \cellcolor{myredfor1} 1.1 (0.2) & \cellcolor{myredfor1}
P30 & \cellcolor{myredfor1} 1.2 (0.4)\\ \hline

\cellcolor{myredfor1}
P11 & \cellcolor{myredfor1} 1.4 (0.6) & \cellcolor{myredfor1}
P31 & \cellcolor{myredfor1} 1.8 (0.9) \\ \hline

\cellcolor{myredfor2}
P12 & \cellcolor{myredfor2} 2.2 (1.0) & \cellcolor{myredfor2}
P32 & \cellcolor{myredfor2} 2.7 (1.1)\\ \hline

\cellcolor{myredfor2}
P13 & \cellcolor{myredfor2} 2.5 (1.5) & \cellcolor{myredfor1}
P33 & \cellcolor{myredfor1} 1.8 (0.8) \\ \hline

\cellcolor{myredfor1}
P14 & \cellcolor{myredfor1} 1.6 (1.0) & \cellcolor{myredfor2}
P34 & \cellcolor{myredfor2} 2.0 (0.8) \\ \hline

\cellcolor{mycyanfor3}
P15 & \cellcolor{mycyanfor3} 3.2 (1.1) & \cellcolor{myredfor1}
P35 & \cellcolor{myredfor1} 1.0 (0.0)\\ \hline

\cellcolor{myredfor1}
P16 & \cellcolor{myredfor1} 1.4 (0.6) & \cellcolor{myredfor2}
P36 & \cellcolor{myredfor2} 2.1 (0.7)\\ \hline

\cellcolor{myredfor1}
P17 & \cellcolor{myredfor1} 1.8 (0.5) & \cellcolor{myredfor2}
P37 & \cellcolor{myredfor2} 2.2 (1.2) \\ \hline

\cellcolor{myredfor2}
P18 & \cellcolor{myredfor2} 2.2 (0.9) & \cellcolor{mycyanfor3}
P39 & \cellcolor{mycyanfor3} 3.1 (1.1) \\ \hline

\cellcolor{myredfor2}
P19 & \cellcolor{myredfor2} 2.2 (0.9) & \cellcolor{myredfor1}
P41 & \cellcolor{myredfor1} 1.0 (0.0) \\ \hline

\cellcolor{myredfor2}
P20 & \cellcolor{myredfor2} 2.0 (0.9) & \cellcolor{myredfor2}
P42 & \cellcolor{myredfor2} 2.6 (1.3) \\ \hline

\cellcolor{myredfor1}
P21 & \cellcolor{myredfor1} 1.8 (0.7) & \cellcolor{myredfor2}
P43 & \cellcolor{myredfor2} 2.0 (0.0) \\ \hline

\cellcolor{myredfor1}
P22 & \cellcolor{myredfor1} 1.0 (0.0) & \cellcolor{myredfor2}
P46 & \cellcolor{myredfor2} 2.9 (1.3) \\ \hlineB{4}

    \end{tabular}}
    \caption{Average sharing tendencies of all participants based on the Likert Scale from 1 - ``Very Unlikely'' to 5 - ``Very Likely''. Standard deviations are given in the parentheses.}
    \label{tab:participant_avg_tendencies_for_emojis}
    \Description{Average sharing tendencies as Likert Score 1 to 5. Table shows that minimum average tendency is 1 with the standard deviation of 0 and maximum average tendency is 4 with the standard deviation of 0. Most participants have an average smaller than 3 which corresponds to "Unsure". Only 4 participants have an average greater than it. }
\end{table}

%Change in Decisions
\subsection{Discrepancy in Expressed Attitudes and Choices:}
\label{sec-discrepancy}

In more than half of these scenarios (360), the participants stated their choice as ``Not Share'' in the first round. This is not surprising considering that the initial tendencies to share were also largely low. However, we also see that their behavior was not always consistent with the stated tendencies. Table \ref{tab:tendencies_and_decision_pairs} shows the distribution of choice pairs and the tendencies. For example, for $57$ cases where a picture was initially set to be "Very Unlikely" to be shared, both the first and second round choices were ``Share''. On the other hand, there were $41$ cases where a picture was initially set to be "Likely" and $5$ cases set to "Very Likely" when the first and second round choices were both "Not Share". We attribute this to the well-known privacy paradox, where the privacy attitudes of individuals sometimes do not match their privacy behavior~\cite{acquisti2015privacy}.

\begin{table}[h!]
    \centering
    \scalebox{0.8}{
    \begin{tabular}{V{4}cV{4}c|c|c|c|cV{4}}
        \hlineB{4}
         \textbf{Choices} & \multicolumn{5}{cV{4}} {\textbf{Tendency to Share}} \\\cline{2-6}\textbf{($\mathbf{1^{st}}$ Round, $\mathbf{2^{nd}}$ Round)}& \textbf{``Very Unlikely''} & \textbf{``Unlikely''} & \textbf{``Unsure''} & \textbf{``Likely''} & \textbf{``Very Likely''}  \\ \hlineB{4}
         (Not Share, Not Share) & 156 & 88 & 54 & 41 & 5 \\ \hline

         (Share, Not Share) & 7 &
 4 & 14 & 9 & 2 \\ \hline
         (Not Share, Share) & 3 & 4 & 4 & 4 & 1 \\ \hline

 (Share, Share) & 57 & 52 & 62 & 63 & 10 \\ \hlineB{4}

          \textbf{Total} & 223 & 148 & 134 & 117 & 18\\ \hlineB{4}
    \end{tabular}}
    \caption{Tendencies and corresponding choice pairs for the $1^{st}$ and $2^{nd}$ round of the scenario instances}
    \label{tab:tendencies_and_decision_pairs}
    \Description{Table shows the distribution of tendencies for the individual images that are used to create the group pictures in the scenario instances compared to the choice pairs for the participants in round 1 and round 2 in the game.
(Not Share, Not Share) L1: 156 L2: 88 L3: 54 L4: 41 L5: 5
(Share, Not Share) L1: 7 L2: 4 L3: 14 L4: 9 L5: 2
(Not Share, Share) L1: 3 L2: 4 L3: 4 L4: 4 L5: 1
(Share, Share) L1: 57 L2: 52 L3: 62 L4: 63 L5: 10
Total L1: 223 L2: 148 L3: 134 L4: 117 L5: 18}
\end{table}

\subsection{User Profiles}
\label{sec-profiles}

We define user profiles based on users' engagement in the game. This engagement can be understood through the changing of choices or bids during the game, from the first round to the second round.  We have first looked at if participants varied in their behavior in changing choices and bids and observed important differences. For example, ``Share'' and ``Not Share'' choices in the first round were kept in the second round in $244$ and $344$ cases, respectively. The participants changed their choice from ``Share'' to ``Not Share'' in $36$ cases while the change from ``Not Share'' to ``Share''
was observed in $16$ cases. The participants won the first round of the game in $453$ cases, where the game was lost in the second round in $18$ cases. For $71$ cases, the participants won the game in the second round, while in $116$ cases the game was lost in both rounds. These demonstrate that participants could use the system to calibrate their individual decisions as they saw fit but also the game was challenging enough that their choices could not be honored at all times.

\begin{table}[th]
    \centering
    \scalebox{0.83}{\begin{tabular}{|c|p{50mm}|p{50mm}|c|}
    \hline
    \textbf{Profile} & \textbf{Description} & \textbf{Set of Participants} & \textbf{\# of Participants} \\
    \hline
    Profile 1 & \textit{never changes choice/bid}  & \{ P35, P42 \} & 2 \\ \hline
    Profile 2 & \textit{changes choice but never changes bid} & \{ P6, P26, P28, P31 \} & 4 \\ \hline
    Profile 3 & \textit{changes bid but never changes choice} & \{ P1, P10, P11, P16, P17, P21, P22, P23, P27, P30, P32, P33, P43, P46 \} & 14\\ \hline
    Profile 4 & \textit{changes  choice and/or bid} & \{ P2, P3, P4, P5, P8, P12, P13, P14, P15, P18, P19, P20, P24, P25, P29, P34, P36, P37, P39, P41 \} & 20 \\ \hline
    \end{tabular}}
    \caption{User profiles based on their interactions with the game}
    \label{tab:participant_profiles}
    \Description{Information about user profiles:

Profile 1 - never changes choice/bid - 2 participants as P35, P42;
Profile 2 - changes choice but never changes bid - 4 participants as P6, P26, P28, P31;
Profile 3 - changes bid but never changes choice - 15 participants as P1, P10, P11, P16, P17, P21, P22,
P23, P27, P30, P32, P33, P43, P46;
Profile 4 - occasionally changing either choice
and/or bid - 20 participants as P2, P3, P4, P5, P8, P12, P13, P14, P15,
P18, P19, P20, P24, P25, P29, P34, P36,
P37, P39, P41}
\end{table}

\begin{table}[t!]
    \centering
    \scalebox{0.75}{
    \begin{tabular}{|c|c|c|c|c|c|}
    \hline
        \multirow{3}{*}{\textbf{Profile}}  &
        \textbf{Average of } & \textbf{Average }
        & \textbf{Average Number of}
        & \textbf{Average Number of} & \textbf{Average Number of} \\
        &
        \textbf{Average Tendencies} & \textbf{Number of }  & \textbf{Choices as ``Share''} & \textbf{Choices as ``Share''} & \textbf{Changes in}  \\
        &  \textbf{to Share} & \textbf{Diversions} & \textbf{in $1^{st}$ round} &  \textbf{in $2^{nd}$ round} & \textbf{Choice or Bid} \\ \hline

        Profile 1 & 1.8 (1.1) & 6.0 (2.8) & 8.5 (0.7) & 8.5 (0.7) & 0.0 (0.0) \\ \hline
        Profile 2 & 1.9 (0.7) & 6.5 (1.0) & 5.5 (1.3) & 4.8 (2.2) & 2.3 (1.6) \\ \hline
        Profile 3 & 1.8 (0.6) & 3.1 (2.8) & 4.9 (3.3) & 4.9 (3.3) & 7.4 (4.0) \\ \hline
        Profile 4 & 2.2 (0.6) & 4.5 (2.6) & 8.7 (2.3) & 7.8 (2.4) & 7.8 (4.0) \\ \hline

    \end{tabular}}
    \caption{Statistics for profiles in terms of average tendencies to share, number of diversions, choices as ``Share'' in both rounds of the game. Standard deviations for the averages are given in the parentheses.}
    \label{tab:profiles_statistics}
    \Description{Statistics about profiles:

Profile 1 - Average of Average Tendencies to Share: 1.8 (with standard deviation of 1.1) - Average Number of Divisions: 6.0 (with standard deviation of 2.8) - Average Number of Choices in First Round as 'Share': 8.5 (with standard deviation of 0.7) - Average Number of Choices in Second Round as 'Share': 8.5 (with standard deviation of 0.7) - Average Number of Changes in Choice or Bid: 0.0 (with standard deviation of 0.0)
Profile 2 - Average of Average Tendencies to Share: 1.9 (with standard deviation of 0.7) - Average Number of Divisions: 6.5 (with standard deviation of 1.0) - Average Number of Choices in First Round as 'Share': 5.5 (with standard deviation of 1.3) - Average Number of Final Choices as 'Share': 4.8 (with standard deviation of 2.2) - Average Number of Changes in Choice or Bid: 2.3 (with standard deviation of 1.6)
Profile 3 - Average of Average Tendencies to Share: 2.0 (with standard deviation of 0.8) - Average Number of Divisions: 3.3 (with standard deviation of 2.8) - Average Number of Choices in First Round as 'Share': 5.2 (with standard deviation of 3.4) - Average Number of Choices in Second Round as 'Share': 5.2 (with standard deviation of 3.4) - Average Number of Changes in Choice or Bid: 7.4 (with standard deviation of 4.0)
Profile 4 - Average of Average Tendencies to Share: 2.2 (with standard deviation of 0.6) - Average Number of Divisions: 4.5 (with standard deviation of 2.6) - Average Number of Choices in First Round as 'Share': 8.7 (with standard deviation of 2.3) - Average Number of Choices in Second Round as 'Share': 7.8 (with standard deviation of 2.4) - Average Number of Changes in Choice or Bid: 7.8 (with standard deviation of 4.0)}
\end{table}

An immediate question is whether we can categorize this information into user profiles based on whether they change their choice and/or bid during the game.

\textbf{\textit{Observation 1.1:}}
Table \ref{tab:participant_profiles} depicts our categorization, with their descriptions, the set of participants who fit into that profile and the cardinality of that set. As can be seen from this table, Profile 1 denotes users who do not change their choices or bids at all during the entire game. There are two participants who fall into this profile.  Profile 2 denotes users that are somehow more flexible regarding their privacy choice but do not consider to change their bids through the game. Profile 3 denotes users who are certain on the choice but do vary their bids based on the interactions in the game. There are 14 participants, who fall into this profile.  Profile 4 denotes users who do change their choice and bid at various times in the game. There are 20 such participants.

We associate bid changing as a behavior of engaging with the game as it requires a thorough thinking of consequences. We then study to what extent participate exhibit engaging behavior. Since participants in Profiles 3 and 4 exhibit this behavior, we consider them together to analyze this behavior. We calculate the average number of changes in choice or bid for the participants in both profiles as 7.6 with a standard deviation of 4.0. The same calculation for the remaining participants leads to the average value of 1.5 with a standard deviation of 1.7. The difference between these two groups shows a significant distinction in their characteristics (The effect size to compare the groups was determined with \textit{Cohen's d} with \textit{d} > 0.8).

\textbf{\textit{Finding 1.1}:} As most participants fit into either Profile 3 and Profile 4 by occasionally changing their choices or bids in almost half of the scenarios, we claim that most participants want to play an active role in the group-decision of privacy by engaging with game.

\textbf{\textit{Observation 1.2}:} Table \ref{tab:profiles_statistics} shows the statistics for the profiles in terms of average tendencies to share, number of diversions (preferences stated in first rounds which contradict with declared tendency), choices as ``Share'' in both first and second rounds of the cases. Average tendency to share increases from Profile 1 to 4, yet there is no significant difference. Diversions have been less observed in Profile 3, with the participants who occasionally change their bid but always keep their choice. Profile 3 is also the group that has the smallest average number of ``Share'' choices for the first round of the cases. Profile 4, on the other hand, is the second profile that the diversions have been less observed. If we again merge these two profiles, we calculate the average number of diversions for engaged participants as 3.9 with the standard deviation of 2.6. The same calculation for the remaining participants leads to the average value of 6.3 with a standard deviation of 1.3. This shows a significant distinction for the characteristic of engaged users to keep their individual preferences in making the group-decision (\textit{Cohen's d} > 0.8).

\textbf{\textit{Finding 1.2}:} The results presented above indicate that all profiles somehow divert from their initial tendencies to share. This leads us to think the understanding of users for their privacy in the individual photos and the group photos are not always same. While engaged users are less likely to divert from their initial tendencies to share, Profile 3, as the participants who tend to be more private and play the game more effectively by adjusting their bids divert from their initial tendencies the least.

\textbf{\textit{Observation 1.3}:}  Using Table \ref{tab:profiles_statistics}, we compare the average numbers of ``Share'' choices in the first and second rounds of the game. This comparison gives us the direction in which preferences of participants mostly shifted in the second round. Obviously, these averages will not change for Profile 1 and 3 for the second rounds, yet we observe a decrease for both Profile 2 and 4.

\textbf{\textit{Finding 1.3}:} Based on the decrease in the averages for ``Share'' choices in Profile 2 and 4, we claim that most users are less flexible while changing their choice from ``Not Share'' to ``Share'' than in the opposite case. This finding increases our motivation to consider the influence of the users' privacy choices in their behaviors, which is also in line with the control theory of privacy \cite{schoeman1984philosophical}.

\section{Behaviors in Multi-Party Privacy Decision Making}\label{sec:behaviors_in_multiparty_privacy}

Next we examine the behaviors that build up the user profiles in detail, particularly how users react to a group-decision in certain cases, and the motivation behind these reactions so that we can answer the research questions RQ2 and RQ3 in Section \ref{sec:keeping_privacy_choice_related_to_rq2} and Section \ref{sec:changing_privacy_choice_related_to_rq3} respectively.

In order to understand the various behaviors, we study when participants change their privacy choice and how or if they adjust their bid. We start by analyzing the correlation between the change of privacy choice and several conditions that a user can encounter in the game. In order to measure this correlation, we use Pearson's correlation. One alternative approach could be to use binary logistic regression to find important factors behind the change in privacy choice. However, the said factors in the game are not independent of each other, which makes the use of binary logistic regression inapplicable for our analysis.

We want to understand if one changes the privacy choice (1) if they already have a favorable outcome, (2) if their choice is not supported by the majority, (3) if they did not want to share it in the first round, and (4) if they initially identified their likeliness of sharing low. Table \ref{tab:pearson_s_r} provides these four conditions and their correlations for the cases where the participants change their privacy choices. There is no strong correlation for any condition (|r-value| < .4 for all), but they are all statistically significant (p-value < .05) to consider their influence in the change of users' preferences. This suggests that such a change can stem from not simply one condition but many different conditions all together in certain cases, which also indicates the complexity of the group-decisions for privacy. Nevertheless, the outcome in the first round of the game seems to have the greatest influence on the decision whether the user is going to keep or change their privacy choice in the second round.

\begin{figure}[H]
    \includegraphics[scale=0.84]{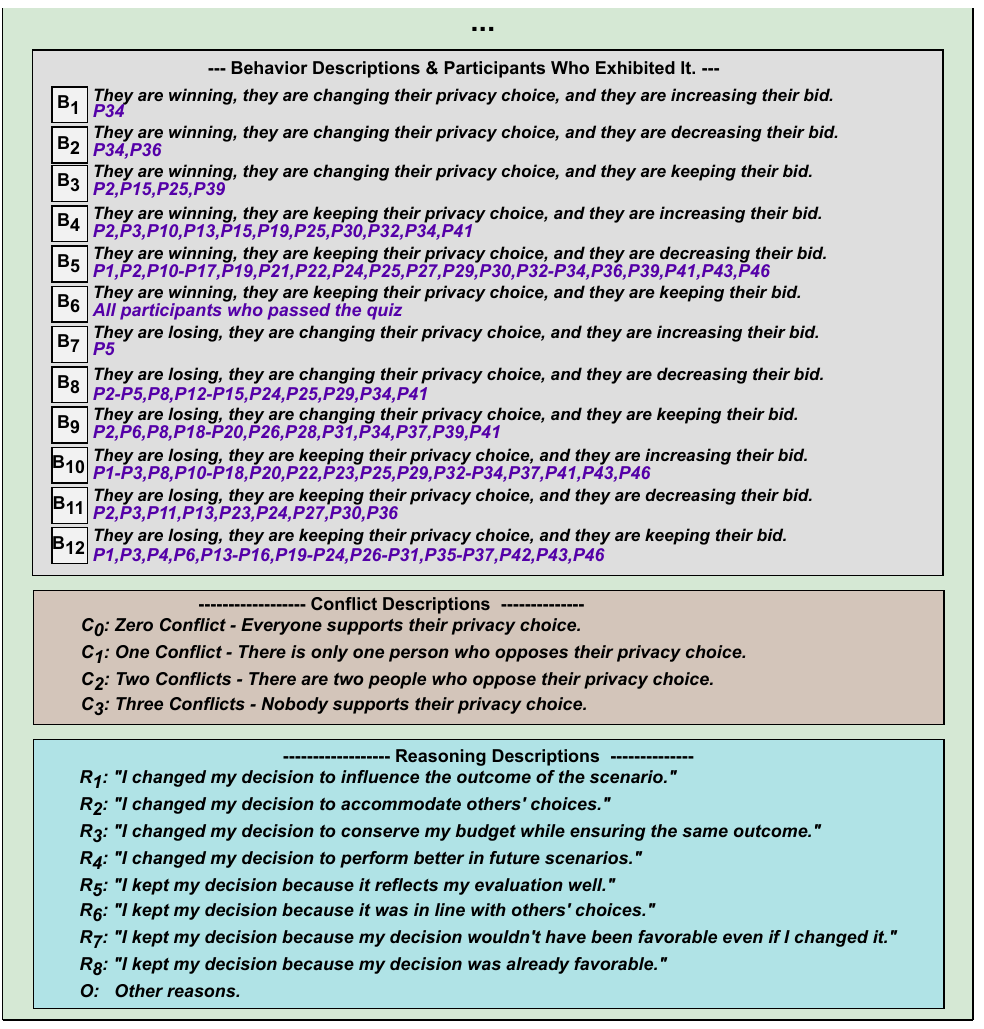}
    \caption{Overview of the results and details of RESOLVE}
    \label{fig:all_behaviors_p2}
    \Description{Second part of the Fig. 6 showing the explanation for each behavior:
    B1) They are winning, they are changing their privacy choice, and they are increasing their bid.
    B2) They are winning, they are changing their privacy choice, and they are decreasing their bid.
    B3) They are winning, they are changing their privacy choice, and they are keeping their bid.
    B4) They are winning, they are keeping their privacy choice, and they are increasing their bid.
    B5) They are winning, they are keeping their privacy choice, and they are decreasing their bid.
    B6) They are winning, they are keeping their privacy choice and their bid.
    B7) They are losing, they are changing their privacy choice, and they are increasing their bid.
    B8) They are losing, they are changing their privacy choice, and they are decreasing their bid.
    B9) They are losing, they are changing their privacy choice, and they are keeping their bid.
    B10) They are losing, they are keeping their privacy choice, and they are increasing their bid.
    B11) They are losing, they are keeping their privacy choice, and they are decreasing their bid.
    B12) They are losing, they are keeping their privacy choice and their bid.

    For each behavior the participants who displayed it is also listed. Details for significant ones are provided in the corresponding figures.

    Conflict descriptions in the group:

    C0)  Zero Conflict - Everyone supports their privacy choice.
    C1)  One Conflict - There is only one person who opposes their privacy choice.
    C2) Two Conflicts - There are two people who oppose their privacy choice.
    C3) Three Conflicts - Nobody supports their privacy choice.

    Reasoning Descriptions:

    R1) "I changed my decision to influence the outcome of the scenario."
    R2) "I changed my decision to accommodate others' choices."
    R3) "I changed my decision to conserve my budget while ensuring the same outcome."
    R4)  "I changed my decision to perform better in future scenarios.
    R5) "I kept my decision because it reflects my evaluation well."
    R6) "I kept my decision because it was in line with others' choices."
    R7)  "I kept my decision because my decision wouldn't have been favorable even if I changed it."
    R8) "I kept my decision because my decision was already favorable."
    O) Other reasons.}
\end{figure}

Following up on this idea, we create a hierarchy of behaviors based on the outcome in the first round, the possible change in both the privacy choice and the bid amount of the participant in first round.
Figure \ref{fig:all_behaviors_p2} shows a comprehensive overview of this hierarchy. In order to ensure its readability, this figure is spread in two pages. Starting with the box, "Are they winning?", different actions of the participants can be traced into different boxes where we observe occurrence of 12 different behaviors that are labeled as $B_1$-$B_{12}$ and described in the figure in detail. For example, when the participant loses the first round of the game, we can observe the behavior $B_8$, if they decide to accommodate the group-decision and to decrease their bid to save some credits meanwhile. However, if they are highly motivated for their privacy choice, they can also display $B_{10}$ in the same circumstances by increasing their bid for the same choice to win the second round of the game.

\begin{table}[h]
    \centering
    \scalebox{0.7}{
    \begin{tabular}{|l|c|}
        \hline
        \multirow{2}{*}{\textbf{Condition}} & 
        \textbf{Pearson's Correlation with} \\
        & \textbf{Keeping the Privacy Choice (r-value)} \\ \hline
        Participant gets a favorable outcome in the first round of the game. & \textbf{.375} \\ \hline

        Participant's choice is not supported by the majority of the group. & -.217 \\ \hline
        
        Participant selects ``Not Share'' in the first round of the game. & .153 \\ \hline
        
        Participant initially tend to ``Not Share'' based on Likert score. & .141 \\ \hline
        
        \end{tabular}}
    \caption{Pearson's Correlations for different conditions in scenarios compared to keeping the privacy choice. All of the listed correlations are weak (|r-value| < .4) but significant with p-value < .05.}
    \label{tab:pearson_s_r}
    \Description{Pearson's r value for different factors in scenarios compared to keeping the privacy choice. The strongest correlation is with the condition of ``Participant gets a favorable outcome in the first round of the game.'' with the r-value of .375. Remaining correlations are listed below:
    The condition: ``Participant's choice is not supported by the majority of the group.'' - r-value: -.217.
    The condition: ``Participant selects ``Not Share'' in the first round of the game.'' - r-value: .153.
    The condition: ``Participant initially tend to ``Not Share'' based on Likert score.'' - r-value: .141.
    All correlations are weak with the absolute values of r being less than .4 but also significant with p-values being less than .05.}
\end{table}

\begin{figure}[H]
    \includegraphics[scale=0.84]{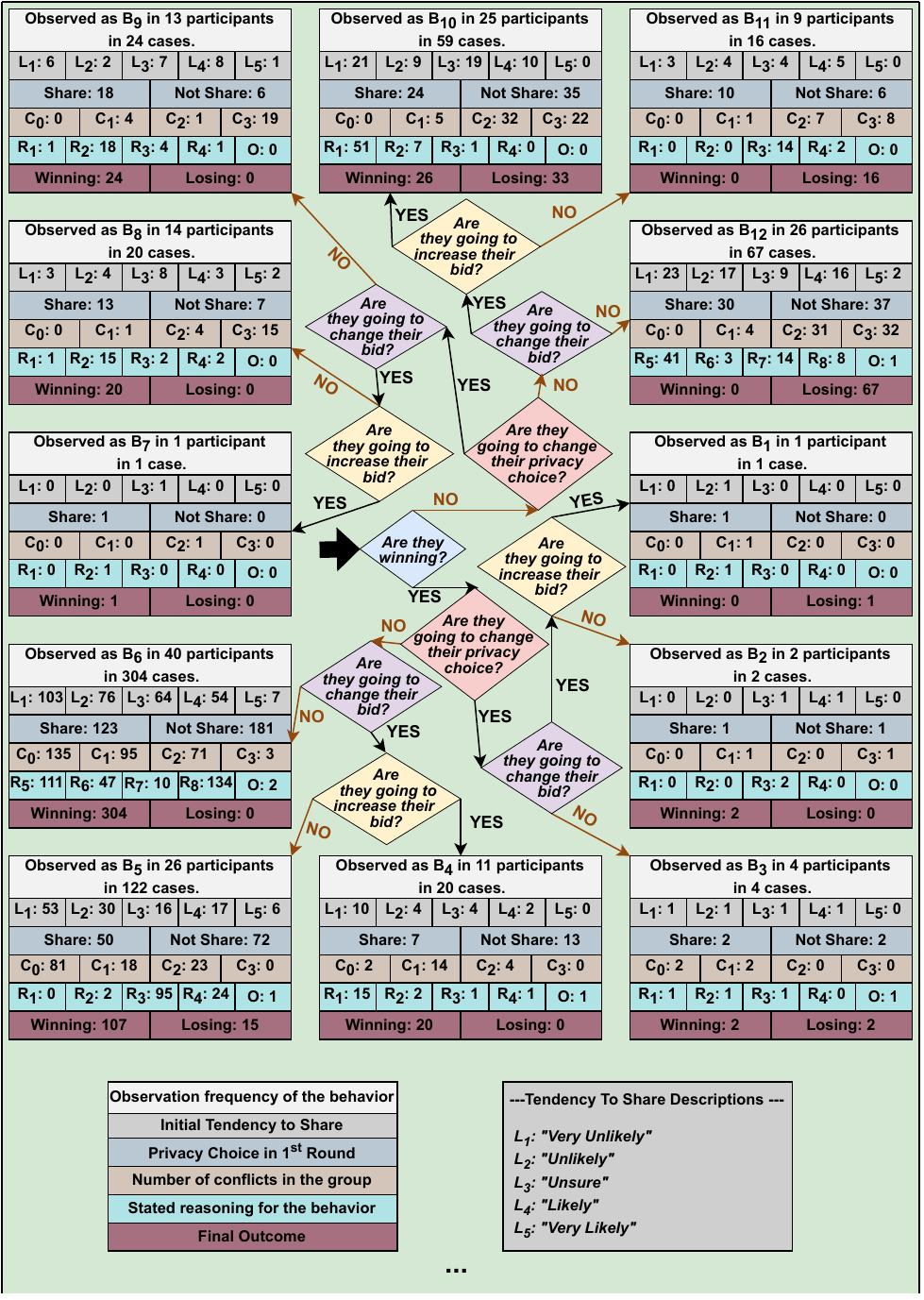}
    \label{fig:all_behaviors}
    \Description{First part of the Fig. 6 showing the overview of the results. Starting from the "Are they winning the box?" decision box, it follows to the two decision boxes with the questions of "Are they going to change their privacy choice", and "Are they going to change their bid?". If the answer for the last question is yes, then a decision box questions "Are they going to increase their bid?". Each decision boxes lead a binary selection "Yes" or "No". Overall the decisions lead to twelve behaviors where for each detailed statistics are presented. Significant behaviors are examined in detail in the paper itself.}
\end{figure}

We particularly study the distribution of occurrences for all of these behaviors including: i) how many different participants displayed the behavior in how many cases in total, ii) the distribution of the participants' likeliness to share the content at the time of taking the individual photo (abbreviated as $L_1$-$L_5$), iii) the distribution of the choices in the first round as ``Share'' or ``Not Share'', iv) the number of conflicts in the group that the participant faced with based on their choice in the first round, \textit{which also indicates the number of other users who lost the first round of the game} (abbreviated as $C_0$-$C_3$), v) the distribution of possible reasons that participants can choose to describe their behavior regarding the possible change in their individual decisions including their privacy choice and their bid (labeled as $R_1$-$R_8$), and vi) definite outcomes of cases after the second round for the participants displaying the behavior. In this way, each behavior can also be related to the factors presented in Table \ref{tab:pearson_s_r}.

What the abbreviations above ($L_1$, $C_0$ etc.) stand for are also listed in Figure \ref{fig:all_behaviors_p2} along with the complete phrases that are presented to participants to state the possible reasons behind their behaviors as $R_1$ to $R_8$. As described in this figure, behaviors are first separated according to the outcome of the scenario in the participants' point of view. Hence, the first six behaviors are observed when the participant wins the game in the first round of a given case, in other words, when they get a favorable outcome in the first run. The rest of the behaviors are the ones that the participant can display while not getting a favorable outcome in the first round of a game. The second criterion to categorize the behaviors is the possible change in privacy choice of the participant. The third and the fourth decision criteria are about the bid that the participant used to support their choice. While the first determines whether the participant is going to change their bid, and if this is so, the last criterion checks whether the final bid of the participant is greater or less than the initial bid.

Table \ref{tab:behavior_reasoning_correlations} shows the behaviors that are significantly observed in the participants along with participant profiles that are mostly associated with these behaviors and finally, the reasons that participants mostly selected to state their motivations behind these behaviors. While trying to answer our remaining research questions in the following subsections we will refer these statistics as well.

\begin{table}[H]
    \centering
    \begin{tabular}{|c|c|c|}
       \hline  \textbf{Behavior}
       & \textbf{Profile(s)}& \textbf{Reason(s)} \\ \hline
       $B_5$ & Profile 3 (r-value: .20) & $R_3$ (r-value: .84), $R_4$ (r-value: .38) \\ \hline
       $B_6$ & Profile 1 (r-value: .14), Profile 2 (r-value: .45) & $R_5$ (r-value: .39), $R_8$  (r-value: .45) \\ \hline
       $B_8$ & Profile 4 (r-value: .45)& \multirow{2}{*}{$R_2$ (r-value: .73)} \\ \cline{1-2}
       $B_9$ & Profile 2 (r-value: .47) & \\ \hline
       $B_{10}$ & Profile 4 (r-value: .19)  & $R_1$ (r-value: .89) \\ \hline
       $B_{12}$ & Profile 1 (r-value: .35), Profile 2 (r-value: .29) & $R_5$ (r-value: .67), $R_7$ (r-value: .35) \\ \hline
    \end{tabular}
    \caption{Significant behaviors and positively correlated profiles and reasons with them. All correlations are significant (p-value < .05)}
    \label{tab:behavior_reasoning_correlations}
    \Description{This table presents the significant behaviors and their most associated (positively correlated) profiles and reasons that the participant used to state their motivations behind those behaviors. Associations are calculated using Pearson's Correlation (r-value).
    B5 is mostly associated with Profile 3 (r-value: .20) and Reasons R3 (r-value: .84) and R4 (r-value: .38).
    B6 is mostly associated with Profile 1 (r-value: .14) and Profile 2 (r-value: .45) and Reasons R5 (r-value: .39) and R8 (r-value: .45).
    B8 and B9 are mostly associated with Reason R2 (r-value: .73) while the former is mostly associated with Profile 4 (r-value: .45), the latter is mostly associated with Profile 2 (r-value: .47).
    B10 is mostly associated with Profile 4 (r-value: .19) and Reason R1 (r-value: .89).
    B12 is mostly associated with Profile 1 (r-value: .35) and Profile 2 (r-value: .29) and Reasons R5 (r-value: .67) and R7 (r-value: .35).
    }
\end{table}

\subsection{Upholding The Privacy Preferences in Opposition of The Majority}\label{sec:keeping_privacy_choice_related_to_rq2}

Here, we look at the cases where the users do not change their privacy choices. In particular, we are interested in such cases where the majority of the group does not support the user's choice. We observe $B_4$ to $B_6$, when the user wins the first round and decides to keep their privacy choice. $B_{10}$ to $B_{12}$ are observed in the opposite case; that is, when the user loses the first round of the game and yet does not change their privacy choice to accommodate the group-decision. In both cases, these behaviors can be interpreted as somehow self-oriented rather than altruistic. In the remaining behaviors, $B_1$ to $B_3$ and $B_7$ to $B_9$, the users change their choice after the first round.

\textbf{\textit{Observation 2.1}:} We observe that in $453$ out of $640$ cases, the participants obtain a favorable outcome in the first run of the game. Only in $7$ cases (exhibited by $B_1$ to $B_3$) they change their privacy choice. In the remaining $446$ cases (exhibited by $B_4$ to $B_6$), they do not change their privacy choice in the second round. It is interesting to look at if the disagreement with others' choices plays a role in these results. By looking at the counts in the $C_0$ (no conflict) and $C_1$ (only one conflict) fields, we can see that in most of these scenario instances the majority of the group agree with the participant's choice: no conflict in total 218 cases for behaviors $B_4$ to $B_6$ and only one different opinion in total 127 cases. Hence, there might not have been a reason for the participant to change their choice. However, we also observe that even when the participant's choice was not supported by the majority of the group ($C_2$ and $C_3$ totaling 101 such cases), the participants keep their privacy choice if they win the first round of the game. It is also worth mentioning only 6 participants changed their privacy choice in such cases, while all participants kept their privacy choice while they got the favorable outcome at least once.

To further investigate the significance of the correlation we found earlier between the favorable outcomes that the participant gets in the first round and the change in their privacy choice, we apply t-test. While the null hypothesis states that there is no relation between the outcome and the change in the privacy choice, the alternative hypothesis suggests that users do not change their privacy choice when they win the game in first round. To reject the null hypothesis, t-value (\(r\sqrt{df / (1 - r^2)}\)) must be greater than its critical value (1.686) for given degree of freedom ($df$ = $40 - 2$) and confidence interval ($\alpha = .95$). When we place r-value as $.375$ into the formula, we get t-value as 2.49 which is sufficient to reject the null hypothesis. This proves that the correlation is significant enough to accept that \textit{users do not change their privacy choices when they get a favorable outcome in the first round of the game.}

\textbf{\textit{Finding 2.1}}: When the users are winning the game, they do not change their choice even if it contradicts with the choices of the majority in the group. A few counterexamples for this inference are only observed with Profile 4 as the set of participants who have relatively less concern about their privacy and more tendency to share the contents initially. Yet, for the remaining observations, it is interesting to see that when a user is sufficiently motivated for their privacy choice to win the game, they do not alter it for the benefit of others.

Although other factors do not present a sufficiently strong correlation to pass the hypothesis test and produce a general rule from the results, we will use them in the analysis of remaining observations to give the likelihood of behaviors in certain conditions for our participants.

\textbf{\textit{Observation 2.2}:} The users largely keep their choices in the second round if they are winning; but do they change their bids? To answer this, we will first have a look at the winning cases where the participants kept their choice as ``Share''. $17$ different participants in $50$ cases decreased their bid in the second round, whereas more than half of the participants (34) kept their initial bids in 123 cases. These two circumstances lead to the behaviors labeled as $B_5$ and $B_6$ respectively.

Next, we look at this when the participants’ choice is ``Not Share'' in the first round. This time $B_5$ was displayed by $23$ different participants who wanted to keep the content private in $72$ cases, $39$ participants chose $B_6$ by keeping their initial bid along with their privacy choice as  ``Not Share'' in $181$ cases. While $23$ participants occasionally changed their bid, only one participant always decreased it.

As can be seen in Table \ref{tab:behavior_reasoning_correlations}, the participant who exhibited $B_5$ stated their main reason for this behavior mostly with the $R_3$ (r-value: .84) and $R_4$ (r-value: .38) which denote the aim for saving some credits in the budget and the consideration of performance for the future scenarios respectively. On the other hand, the participants whose behavior was $B_6$ in such cases selected $R_5$ (r-value: .39) and $R_8$ (r-value:.45) for their main motivation, which mean the satisfaction of the participants with their assessment for the scenario and their choice being already favorable respectively. $B_5$ is mostly associated with Profile 3 (r-value: .20) while $B_6$ is mostly associated with Profile 1 (r-value: .14) and Profile 2 (r-value: .45).

When we compare the field $C_0$ for both behaviors with participants who support either share or not share decision,
we can see $B_6$ has been observed in the scenarios where the participants kept their bid by having relatively less support from the group. Yet, the participants displayed $B_5$ by decreasing their bids in the cases where every other member was aligned with the participant’s choice (see Fig. \ref{fig:all_behaviors_p2}).

\textbf{ \textit{Finding 2.2}:}
When winning the game, participants who wanted ``Share'' outcome, sometimes strategically decreased their bids to save credits for future. The motivations that were declared by the participants are consistent with this insight. This was essentially a risky move as by decreasing their bid, the outcome of the auction might have changed to their disadvantage. When the participants' choice was ``Not Share'', the participants were less likely to take risks as they did not decrease their bids when winning as much. Again, this claim is supported by both the motivations indicated by the participants and the fact that almost all participants kept their initial bids despite the complete or partial support from the other members in the group.

\textbf{\textit{Observation 2.3}}: How do the users behave when they are losing the game? We inspect the significant behaviors $B_{10}$ and $B_{12}$, where the participant fights for their privacy by keeping their privacy choice even though they are losing.  When the privacy choice in the first round was "Share", $B_{10}$ was observed in $15$ participants who increased their bid in the second round of $24$ cases. By keeping their initial bids along with the privacy choice as ``Share'', 18 different participants displayed $B_{12}$ in 30 cases in total.

Now we examine these behaviors $B_{10}$ and $B_{12}$ for the cases where the participants preferred not sharing the content. $19$ participants displayed $B_{10}$ by increasing their bid in $35$ to change the outcome in the favor of keeping the content private. $20$ participants decided to participate in the second round of $37$ such cases with the same bid, which led them displaying the behavior $B_{12}$.

For both choices, the participants who displayed $B_{10}$ increased their bids mostly based on the motivation denoted with $R_1$ stating that they intended to change the outcome in the favor of their preferences (r-value: .89). The reasons behind $B_{12}$ on the other hand, were mostly expressed with $R_5$ (r-value: .67) and $R_7$ (r-value: .35). While the former states that their satisfaction for their assessment for the scenario, the second one means that they did not think that they would not be able to change the outcome. $B_{10}$ is mostly associated with Profile 4 (r-value: .19) while $B_{12}$ is mostly associated with Profile 1 (r-value: .35) and Profile 2 (r-value: .29).

The number of participants who always kept their bids
the choice as ``Not Share'' is slightly higher, 14 participants out of 20, compared to the cases with 12 participants having the opposite choice.

Since we are examining the cases where the participants lost the first round of the game, it is natural to see less support from the group for the participants' privacy choices. Yet we observe the participants 9 games in total despite the agreement of the majority to share the content. On the other hand, the participants never lost a game unless there were at least two members disagreeing with the participant and supporting the group-decision to make the content public. Furthermore, in 17 cases out of 38, the participants increased their bids to change the outcome in their favor, although they were left alone regarding their choice for not sharing the content.

\textbf{\textit{Finding 2.3}}:
When losing the game, participants only occasionally fought to "Share" the content but more likely when their choice was "Not Share".
When we compare the group dynamics indicated with $C_0$-$C_3$ fields of $B_{10}$ for both privacy choices, we can see that participants, who wanted to keep the content private, increased their bids to change the outcome despite the lack of support in the group. Furthermore, it is also worth mentioning that if at least one member supports ``Not Share'' choice, these participants never lost a game. That is only possible with greater bids reflecting their high motivation to avoid sharing the content with an audience.

\subsection{Circumstances That Lead to Change in Privacy Preferences}\label{sec:changing_privacy_choice_related_to_rq3}

In this section, we focus on the circumstances that lead people to change their privacy choices. Understanding these circumstances is important to assess the flexibility of users in terms of their privacy choice. Does such a flexibility stem from the respect to others' privacy preferences? Or do they want not to be left alone by their peers? Finding answers to these questions can lead to better decision mechanisms for the privacy of groups.

To try to answer these questions, we particularly look for the behaviors $B_8$ and $B_9$ where the participant loses the game and decides to accommodate the group-decision by giving up their privacy choice in the first round. While the former behavior means that they try to save some credits in such a case by decreasing their bid, the latter is observed when they keep their bid the same, although it supports their new privacy choice in the second round.

\textbf{\textit{Observation 3.1}:} Here we consider the change in the participants' bids for the cases that they lost and changed their privacy choice from ``Share'' to ``Not Share'' in the second round. Since the participants tended to keep their choice even if they lost the game, we observed these behaviors among less than half of the participants. $B_8$ was observed in 10 participants for 13 scenarios who decreased their initial bid while accommodating the group choice to keep the content private. 11 participants displayed $B_9$ by keeping it same for 18 such cases in total.

Most probably the change in privacy choice was less seen in the mechanism, there was no significant correlation between these behaviors and any of the reasons. Yet, as such a change is more critical than the readjustment of the bid, we merged these cases and found out that the cases associated with these behaviors were mostly correlated with $R_2$ (r-value: .73) as can be seen in Table \ref{tab:behavior_reasoning_correlations}. $B_8$ is mostly associated with Profile 4 (r-value: .45) while $B_9$ is mostly associated with Profile 2 (r-value: .47). We observed 9 cases where the participants lost the game even when they got support from one or two members in the group for their choice to share the co-owned content.

\textbf{\textit{Finding 3.1}}: As reflected with $B_8$, participants occasionally saved some credits while accommodating the group choice as ``Not Share''. There were some cases that they lost despite some partial support from the group, which tells us they did not bid in an amount sufficient to dominate the choice for ``Not Share'' in the first round, and they changed their choice instead of trying to win the game in the second round by increasing their bid and supporting ``Share'' decision more. Hence, this observation implies that they were not highly motivated to share the content and they might be more flexible while changing their choice from ``Share'' to ``Not Share''.

\textbf{\textit{Observation 3.2}:}
Finally, we look at the participants displaying $B_8$ and $B_9$ by losing the game in the first round and changing their privacy choice from ``Not Share'' to ``Share''. These behaviors were observed fewer times (13 in total), relative to the cases where the participants preferred sharing the content in the first round and they lost. $7$ participants displayed $B_8$ in $7$ such cases, and only $4$ participants displayed $B_9$ in $6$ these cases.

The distribution of conflicts for both behaviors in this case is $C_0$: 0, $C_1$: 0, $C_2$: 1, $C_3$: 12. We can see that in almost all of them, the participants changed their privacy choice from ``Not Share'' to ``Share'' if they were the only member of the group who wanted to keep the content private.

\textbf{\textit{Finding 3.2}}: The participants who did not want to share the content, were less likely to accommodate the group unless they were left alone. The field $C_1$ is zero for both behaviors in this type of scenario instances. Hence, we can say that if they wanted to keep the content within the group, they did not accommodate if there is only one person in the group who is willing to share. We also observe only one scenario where there is just one member supporting the participant's choice as ``Not Share'' yet they lost the game. Hence, it can be inferred that they {had a greater chance} to win the game if there is at least one member aligned with the choice as ``Not Share'' and they might be convinced to share the content in the second round, only if they believe they cannot change the outcome.

\section{Validity and Limitations of The Study}
\label{sec:validity_and_limitations}

Five different categories of questions are included in the quiz as summarized in Table \ref{tab:quiz_categories}. For each wrong answer of the participant an immediate feedback is provided with a pop-up window, before the participants move to the next one. Hence, the second question in each category gives another chance to the participants to practice in the same context after they learn from their mistakes. This ensured that participants who passed the quiz actually understood all aspects of the game.

\begin{table}[h!]
    \centering
    \begin{tabular}{|c|l|c|}
        \hline
         \textbf{Category} & \textbf{Explanation} & \textbf{Questions} \\ \hline
         1 & Basic Interaction to Express Opinion & 1, 2 \\ \hline
         2 & Determination of the Outcome & 3, 4 \\ \hline
         3 & Payment Calculation for
 a Favorable Outcome & 5, 6 \\ \hline
         4 & Payment Calculation for an Unfavorable Outcome & 7, 8 \\ \hline
         5 & Evaluation of the Performances & 9, 10 \\ \hline
    \end{tabular}
    \caption{Categories in the Quiz}
    \Description{Question Categories in Quiz:

Category 1 - Basic Interaction to Express Opinion - Q1, Q2
Category 2 - Determination of the Outcome - Q3, Q4
Category 3 - Payment Calculation for a Favorable Outcome - Q5, Q6
Category 4 - Payment Calculation for an Unfavorable Outcome - Q7, Q8
Category 5 - Evaluation of the Performances - Q9, Q10}\label{tab:quiz_categories}
\end{table}

Although the participants were trained in an uncontrolled and online manner, the success rate in the quiz was significantly high. While only 5 of the 46 participants failed in the quiz and 3 participants needed 2 or more attempts. 2 participants contacted us to retake the training and reattempt the quiz. One of the failed participants could have reattempted the quiz without contacting us, but did not use this chance. Nevertheless, 38 participants became successful at their first attempt in the quiz.  Overall results of the quiz indicate that even with a limited training experience, most of the users could understand how they express their privacy choices in the game, how the group-decision and their payment for the resulting decision would be determined.

Table \ref{tab:success_rates_each_question} shows the success rates for each question in the quiz. The first question, where participants were given a task to select ``Not Share'' as their choice and give at least 11 credits to support this choice, has the lowest rate among all questions. One possible explanation for this result is that the participants were not careful enough to grasp what the question required and directly stated their actual choice for the given scenario in the beginning of the quiz. This claim is supported by the high success rate of the second question where the same task was given with different privacy choice and bid amount.

\begin{table}[H]
    \centering
    \begin{tabular}{|c|c|c|}
        \hline
         \textbf{Category} & \textbf{Question} & \textbf{Success Rate} \\ \hline
         \multirow{2}{*}{1} & 1 & 65.2\% \\ \cline{2-3}
         & 2 & 91.3\% \\ \hline
         \multirow{2}{*}{2} & 3 & 97.8\% \\ \cline{2-3}
         & 4 & 91.3\% \\ \hline
         \multirow{2}{*}{3} & 5 & 82.6\% \\ \cline{2-3}
         & 6 & 76.1\% \\ \hline
         \multirow{2}{*}{4} & 7 & 76.1\% \\ \cline{2-3}
         & 8 & 91.3\% \\ \hline
         \multirow{2}{*}{5} & 9 & 95.7\% \\ \cline{2-3}
         & 10 & 89.1\% \\ \hline

    \end{tabular}
    \caption{Success Rates for Each Question in the Categories}
    \label{tab:success_rates_each_question}
    \Description{Success Rates in Quiz Questions. Minimum is 65.2\% with Q1, Maximum is 97.9\% with Q3. Overall rates show a sufficient understanding of participants to play the game effectively.}
\end{table}

Despite the recurrent mistakes that participants have made in some certain questions, the majority of the attempts resulted in success for each question. With that in mind, we can claim that the success rates for each question were in line with the overall view of the quiz results.

Although the quiz covers the main essence of the game, it cannot use all the cases within the same outcomes where the numbers of supporting and opposing people in the group differ. Without being forced to take an action accordingly, participants may not consider all of the possible cases while selecting their amount of bid. Hence, their understanding of the game may be superficial and it may not lead to an effective usage for the players all the time.

While the quiz validates the user understanding of the game, three scenarios ($5^{th}$, $10^{th}$ and  $15^{th}$ scenario that the participant has already taken part in) are presented to the participants again as attention checkers to validate their responses during the experiment phase. If a participant does not have the same privacy choices for any of these three scenarios with the choices they had in actual ones, then they are considered as inattentive. Among $41$ participants, only one of them is excluded due to such inconsistencies in their responses.

$40$ participants on the other hand, completed all 16 scenarios and passed this attention check. Hence, they are considered as eligible for the study. The average completion time of these participants for the overall experiment is 37 minutes 18 seconds with the standard deviation of 15 minutes 1 second. Since the participants are given the option to log out to pause the experiment, this calculation is solely based on the time that they are active in the system.

The interface of our user study enables the participants to express their main reason to keep or change their privacy choices and bid amounts in the second of the group-decision. Although these inputs from the user support the labels we used for their behavior in most cases, there are still some inconsistencies with the indicated reason and the way they act. Considering that this is the most common problem in any system with user interaction, the results may be interpreted within a limited certainty.

Finally, as both demographics and tendencies of the participants (see Table \ref{tab:participant_avg_tendencies_for_emojis} and Table \ref{tab:participant_demography} in Appendix \ref{sec:participant_characteristics}) indicate the majority of the participants were highly educated, somehow concerned about their privacy and tended to keep most of the images private. Hence, the users we observed in our study may present more specific patterns of behaviors that could be different than the average user in OSNs.

\section{Conclusion}\label{sec:conclusion}
Overall, our work identifies interesting privacy dynamics of users within groups. We observe that the individual privacy valuations for pictures differ when the picture is embedded in a group picture. At different situations, there are individuals who would respect the preferences of the others and adjust their sharing behavior to fit them. But, in cases where the privacy of the content is important for the individual, they tend to fight to push for their privacy preferences as well.

These results open up interesting directions to pursue. First, they may shed the light on the possible development of different scales for the factors leading the user behaviors in group-decisions like in the study by Rakibul \textit{et al.} \cite{rakibul_vopp} to measure the value of non-users' privacy from the user's perspective. In our current study, the participant sees themselves in a group picture where there is a fixed number of others, who are unknown. Hence, another important direction is to study whether users' behavior changes based on the identity or the number of others in the group picture. This could help us further understand the privacy dynamics based on the group properties. Another important direction is to formalize these findings as requirements in building privacy management modules into collaborative systems, such as online social networks or co-editing systems. This would enable bring us closer to having (multi-party) privacy-respecting collaborative systems.

\section*{Acknowledgments}
This research was funded by \href{https://www.hybrid-intelligence-centre.nl/}{the Hybrid Intelligence Centre}, a 10-year programme funded by the Dutch Ministry of Education, Culture and Science through the Netherlands Organisation for Scientific Research. H{\"u}seyin Ayd{\i}n is supported by \href{https://www.tubitak.gov.tr}{the Scientific and Technological Research Council of Turkey}, through BIDEB 2219 International Postdoctoral Research Scholarship Program.

\bibliographystyle{ACM-Reference-Format}
\bibliography{privacy}

\clearpage

\appendix
\section{Participant Characteristics}\label{sec:participant_characteristics}

\begin{table}[h!]
    \centering
    \scalebox{0.75}{
    \begin{tabular}{|l|c|}
        \hline
         \textbf{Variable} & \textbf{Distribution} \\ \hline
         Gender & female (47.8\%), male (52.2\%) \\ \hline
         Age &  18-22 (6.5\%), 23-30 (45.7\%), 31-40 (28.3\%), 41-50(17.4\%), 51-64 (2.2\%)
         \\ \hline
         
         Education & master's or doctoral degree (69.6\%), associate or bachelor's degree (21.7\%), undergraduate students (8.7\%) \\ \hline
         Employment  & employed full or part-time (84.8\%), looking for work (2.2\%), self-employed (2.2\%), student (10.8\%) \\ \hline
         Priv. concern in OSNs & very unconcerned (4.3\%), unconcerned (6.5\%), neutral (30.4\%), concerned (39.1\%), very concerned (19.6\%)\\ \hline
         Post freq. on OSNs & never (2.2\%), rarely (54.3\%), occasionally (28.3\%), frequently (15.2\%) \\ \hline
         
    \end{tabular}}
    \caption{The demographic information of the participants and their relations with OSNs.}
    \label{tab:participant_demography}
\end{table}

\section{Privacy Concerns \& Sharing Frequencies in the Profiles}\label{sec:concern_and_frequency_for_profiles}

\begin{table}[h]
    \centering
    \scalebox{0.85}{
    \begin{tabular}{|c|c|c|c|c|c|c|}
        \hline
         \multirow{3}{*}{\textbf{Profiles}} & \multicolumn{5}{c|} {\textbf{Declared Level of Privacy Concern}} 
         & \multirow{3}{*}{\textbf{Average Score}} \\\cline{2-6} & \textbf{``Very} & \multirow{2}{*}{\textbf{``Unconcerned''}}
         & \multirow{2}{*}{\textbf{``Neutral''}} & \multirow{2}{*}{\textbf{``Concerned''}}
         & \textbf{``Very} &  \\
         & \textbf{Unconcerned''} & & & & \textbf{Concerned''} & 
            \\ \hline
         Profile 1 & 0 & 0 & 0 & 2 & 0 & 4.0 (0.0) \\ \hline
         Profile 2 & 0 & 0 & 1 & 3 & 0 & 3.8 (0.9) \\ \hline 
         Profile 3 & 1 & 1 & 6 & 2 & 4 & 3.5 (1.2) \\ \hline 
         Profile 4 & 1 & 1 & 5 & 9 & 4 & 3.7 (1.0) \\ \hline

    \end{tabular}}
    \caption{Profiles and the distribution of their declared level of privacy concerns. Standard deviations for the averages are given in the parentheses.}
    \label{tab:profiles_and_concern}
\end{table}

\begin{table}[h]
    \centering
    \scalebox{0.9}{
    \begin{tabular}{|c|c|c|c|c|c|c|}
        \hline
         \multirow{3}{*}{\textbf{Profiles}} & \multicolumn{5}{c|} {\textbf{Declared Level of Sharing Frequency}} 
         & \multirow{3}{*}{\textbf{Average Score}} \\\cline{2-6} & \multirow{2}{*}{\textbf{``Never''}} & \multirow{2}{*}{\textbf{``Rare''}}
         & \multirow{2}{*}{\textbf{``Occasional''}} & \multirow{2}{*}{\textbf{``Frequently''}}
         & \textbf{``Very} &  \\
         & & & & & \textbf{Frequently''} & 
            \\ \hline
         Profile 1 & 0 & 1 & 1 & 0 & 0 & 2.5 (0.7) \\ \hline
         Profile 2 & 0 & 2 & 2 & 0 & 0 & 2.5 (0.6) \\ \hline 
         Profile 3 & 0 & 4 & 9 & 1 & 0 & 2.4 (0.6) \\ \hline 
         Profile 4 & 1 & 10 & 5 & 4 & 0 & 2.6 (0.9) \\ \hline

    \end{tabular}}
    \caption{Profiles and the distribution of their declared level of sharing frequency. Standard deviations for the averages are given in the parentheses.}
    \label{tab:profiles_and_sharing_frequency}
\end{table}

\clearpage

\section{Performances of The Participants}\label{sec:participant_performances}

\begin{table}[h]
    \centering
    \scalebox{0.8}{
    \begin{tabular}{|c|c|c|c|}
    \hline
    \multirow{2}{*}{\textbf{Participant}} & \textbf{Favorable} & \textbf{Remaining} & \multirow{2}{*}{\textbf{Reward}} \\
    
    & 
    \textbf{Outcomes} &  \textbf{Budget} &  \\ \hline
    P1 & 15 & 112 & 8.6 \\ \hline
P2 & 11 & 15 & 5.7 \\ \hline
P3 & 10 & 123 & 6.2 \\ \hline
P4* & 13 & 165 & 8.2 \\ \hline
P5 & 16 & 15 & 8.2 \\ \hline
P6 & 14 & 61 & 7.6 \\ \hline
P8 & 15 & 127 & 8.8 \\ \hline
P10 & 17 & 170 & 10.2 \\ \hline
P11 & 13 & 141 & 7.9 \\ \hline
P12 & 12 & 123 & 7.2 \\ \hline
P13 & 14 & 166 & 8.7 \\ \hline
P14* & 13 & 151 & 8.0 \\ \hline
P15* & 14 & 119 & 8.2 \\ \hline
P16 & 14 & 86 & 7.9 \\ \hline
P17 & 16 & 177 & 9.8 \\ \hline
P18 & 11 & 168 & 7.2 \\ \hline
P19 & 13 & 20 & 6.7 \\ \hline
P20 & 13 & 135 & 7.8 \\ \hline
P21 & 16 & 105 & 9.1 \\ \hline
P22 & 15 & 153 & 9.0 \\ \hline
P23 & 14 & 102 & 8.0 \\ \hline
P24 & 9 & 141 & 5.9 \\ \hline
P25 & 14 & 85 & 7.8 \\ \hline
P26 & 15 & 136 & 8.9 \\ \hline
P27 & 12 & 147 & 7.5 \\ \hline
P28 & 13 & 222 & 8.7 \\ \hline
P29 & 13 & 191 & 8.4 \\ \hline
P30 & 17 & 139 & 9.9 \\ \hline
P31 & 14 & -5 \tablefootnote{This participant used all their credits in the last scenario. Since the outcome has changed with their decision, their final budget seems negative with the applied tax of 5 credits. If that was an earlier scenario, the participant still would have a positive budget with additional 10 credits for each new scenario. Overall reward for this participant is calculated considering the remaining budget as 0.} & 7.0 \\ \hline
P32 & 17 & 11 & 8.6 \\ \hline
P33 & 17 & 23 & 8.7 \\ \hline
P34 & 12 & 124 & 7.2 \\ \hline
P35 & 11 & 185 & 7.3 \\ \hline
P36 & 9 & 225 & 6.8 \\ \hline
P37* & 12 & 98 & 7.0 \\ \hline
P39 & 16 & 7 & 8.1 \\ \hline
P41 & 15 & 151 & 9.0 \\ \hline
P42 & 10 & 180 & 6.8 \\ \hline
P43 & 11 & 219 & 7.7 \\ \hline
P46* & 15 & 24 & 7.7 \\ \hline
    \end{tabular}
    }
    \caption{Performances of the participants regarding the number of favorable outcomes they got, their remaining budget, and their reward calculated using those two. Participants who are indicated with `*' won the lottery for additional €20 voucher.}
    \label{tab:participant_performances}
\end{table}

\clearpage

\section{Initial Sharing Tendencies of Participants}\label{sec:tendencies_all_emojis}

\begin{table}[h]
    \centering
    \scalebox{0.8}{
    \begin{tabular}{ | p{8mm} | c | c | c | c | c | c | c | c | c | c | c | c | c | c| c | c | c |}
        \hline
        \multirow{2}{*}{\textbf{Part.}} & \multicolumn{16}{c|}{\textbf{Sharing Tendencies for Each Emoji}} &  \textbf{Average} \\ \cline{2-17}
        &
        $\mathbf{E_1}$ & $\mathbf{E_2}$ & $\mathbf{E_3}$ & $\mathbf{E_4}$ & $\mathbf{E_5}$ & $\mathbf{E_6}$ & $\mathbf{E_7}$ & $\mathbf{E_8}$ & $\mathbf{E_9}$ & $\mathbf{E_{10}}$ & $\mathbf{E_{11}}$ & $\mathbf{E_{12}}$ & $\mathbf{E_{13}}$ & $\mathbf{E_{14}}$ & $\mathbf{E_{15}}$ & $\mathbf{E_{16}}$ &  \textbf{Score}\\ \hline
        \rowcolor{myredfor2}
P1 & 3 & 2 & 2 & 4 & 3 & 2 & 1 & 3 & 4 & 2 & 3 & 2 & 4 & 2 & 2 & 4 & 2.7 (0.9)\\ \hline
\rowcolor{myredfor1}
P2 & 3 & 1 & 1 & 1 & 1 & 1 & 1 & 1 & 1 & 1 & 1 & 1 & 1 & 1 & 1 & 1 & 1.1 (0.5)\\ \hline
\rowcolor{myredfor1}
P3 & 4 & 1 & 1 & 1 & 1 & 1 & 1 & 2 & 4 & 1 & 3 & 1 & 1 & 1 & 1 & 4 & 1.8 (1.2)\\ \hline
\rowcolor{myredfor2}
P4 & 5 & 1 & 1 & 4 & 2 & 1 & 5 & 4 & 5 & 1 & 5 & 1 & 4 & 1 & 1 & 1 & 2.6 (1.8)\\ \hline
\rowcolor{myredfor2}
P5 & 4 & 1 & 1 & 2 & 1 & 1 & 1 & 3 & 5 & 2 & 2 & 1 & 3 & 1 & 1 & 3 & 2.0 (1.3)\\ \hline
\rowcolor{myredfor2}
P6 & 2 & 2 & 3 & 4 & 2 & 4 & 2 & 4 & 5 & 2 & 2 & 2 & 2 & 2 & 2 & 4 & 2.8 (1.1)\\ \hline
\rowcolor{myredfor2}
P8 & 1 & 2 & 1 & 3 & 1 & 2 & 2 & 3 & 3 & 2 & 2 & 2 & 2 & 2 & 2 & 2 & 2.0 (0.6)\\ \hline
\rowcolor{myredfor1}
P10 & 2 & 1 & 1 & 1 & 1 & 1 & 1 & 1 & 1 & 1 & 1 & 1 & 1 & 1 & 1 & 1 & 1.1 (0.2)\\ \hline
\rowcolor{myredfor1}
P11 & 2 & 1 & 1 & 2 & 1 & 1 & 1 & 1 & 3 & 1 & 2 & 1 & 1 & 2 & 1 & 2 & 1.4 (0.6)\\ \hline
\rowcolor{myredfor2}
P12 & 3 & 2 & 1 & 1 & 2 & 1 & 1 & 3 & 4 & 2 & 3 & 1 & 3 & 2 & 3 & 3 & 2.2 (1.0)\\ \hline
\rowcolor{myredfor2}
P13 & 2 & 4 & 5 & 2 & 1 & 2 & 1 & 4 & 5 & 1 & 2 & 1 & 4 & 1 & 1 & 4 & 2.5 (1.5)\\ \hline
\rowcolor{myredfor1}
P14 & 4 & 1 & 1 & 1 & 2 & 1 & 1 & 4 & 2 & 2 & 1 & 1 & 1 & 1 & 1 & 1 & 1.6 (1.0)\\ \hline
\rowcolor{mycyanfor3}
P15 & 5 & 2 & 3 & 4 & 1 & 4 & 4 & 4 & 4 & 3 & 2 & 3 & 4 & 2 & 3 & 4 & 3.2 (1.1)\\ \hline
\rowcolor{myredfor1}
P16 & 1 & 1 & 1 & 1 & 2 & 1 & 1 & 2 & 3 & 1 & 1 & 1 & 2 & 1 & 2 & 1 & 1.4 (0.6)\\ \hline
\rowcolor{myredfor1}
P17 & 2 & 2 & 1 & 2 & 1 & 2 & 1 & 2 & 3 & 2 & 2 & 2 & 2 & 2 & 1 & 2 & 1.8 (0.5)\\ \hline
\rowcolor{myredfor2}
P18 & 2 & 3 & 2 & 3 & 3 & 1 & 1 & 2 & 4 & 2 & 3 & 2 & 2 & 3 & 1 & 2 & 2.2 (0.9)\\ \hline
\rowcolor{myredfor2}
P19 & 1 & 3 & 1 & 2 & 2 & 2 & 1 & 2 & 4 & 2 & 4 & 2 & 2 & 2 & 3 & 2 & 2.2 (0.9)\\ \hline
\rowcolor{myredfor2}
P20 & 2 & 1 & 1 & 2 & 2 & 3 & 1 & 3 & 3 & 3 & 3 & 1 & 3 & 1 & 1 & 2 & 2.0 (0.9)\\ \hline
\rowcolor{myredfor1}
P21 & 2 & 2 & 1 & 2 & 1 & 1 & 1 & 2 & 3 & 2 & 2 & 3 & 2 & 2 & 1 & 2 & 1.8 (0.7)\\ \hline
\rowcolor{myredfor1}
P22 & 1 & 1 & 1 & 1 & 1 & 1 & 1 & 1 & 1 & 1 & 1 & 1 & 1 & 1 & 1 & 1 & 1.0 (0.0)\\ \hline
\rowcolor{myredfor1}
P23 & 2 & 1 & 1 & 2 & 1 & 1 & 1 & 2 & 3 & 1 & 2 & 1 & 1 & 1 & 1 & 3 & 1.5 (0.7)\\ \hline
\rowcolor{mycyanfor3}
P24 & 4 & 4 & 5 & 3 & 2 & 4 & 3 & 5 & 4 & 4 & 2 & 4 & 1 & 2 & 3 & 4 & 3.4 (1.1)\\ \hline
\rowcolor{myredfor1}
P25 & 2 & 2 & 2 & 2 & 1 & 2 & 1 & 2 & 3 & 2 & 3 & 1 & 2 & 1 & 1 & 3 & 1.9 (0.7)\\ \hline
\rowcolor{myredfor1}
P26 & 1 & 1 & 1 & 1 & 1 & 1 & 1 & 2 & 1 & 1 & 1 & 1 & 2 & 1 & 1 & 1 & 1.1 (0.3)\\ \hline
\rowcolor{myredfor2}
P27 & 2 & 1 & 1 & 4 & 3 & 1 & 1 & 4 & 5 & 1 & 4 & 2 & 3 & 2 & 1 & 3 & 2.4 (1.4)\\ \hline
\rowcolor{myredfor2}
P28 & 2 & 1 & 1 & 1 & 1 & 4 & 1 & 1 & 4 & 3 & 4 & 1 & 1 & 1 & 4 & 2 & 2.0 (1.3)\\ \hline
\rowcolor{myredfor2}
P29 & 4 & 1 & 2 & 3 & 1 & 4 & 2 & 5 & 4 & 4 & 2 & 3 & 2 & 1 & 1 & 2 & 2.6 (1.3)\\ \hline
\rowcolor{myredfor1}
P30 & 1 & 1 & 2 & 1 & 1 & 1 & 1 & 2 & 1 & 2 & 1 & 1 & 2 & 1 & 1 & 1 & 1.2 (0.4)\\ \hline
\rowcolor{myredfor1}
P31 & 2 & 1 & 2 & 1 & 3 & 1 & 2 & 1 & 4 & 2 & 3 & 2 & 1 & 1 & 1 & 1 & 1.8 (0.9)\\ \hline
\rowcolor{myredfor2}
P32 & 4 & 3 & 2 & 2 & 4 & 3 & 2 & 3 & 3 & 4 & 4 & 4 & 2 & 1 & 1 & 1 & 2.7 (1.1)\\ \hline
\rowcolor{myredfor1}
P33 & 3 & 2 & 2 & 3 & 1 & 3 & 1 & 2 & 3 & 2 & 1 & 2 & 1 & 1 & 1 & 1 & 1.8 (0.8)\\ \hline
\rowcolor{myredfor2}
P34 & 2 & 1 & 2 & 3 & 3 & 2 & 1 & 2 & 3 & 3 & 2 & 2 & 1 & 1 & 1 & 3 & 2.0 (0.8)\\ \hline
\rowcolor{myredfor1}
P35 & 1 & 1 & 1 & 1 & 1 & 1 & 1 & 1 & 1 & 1 & 1 & 1 & 1 & 1 & 1 & 1 & 1.0 (0.0)\\ \hline
\rowcolor{myredfor2}
P36 & 4 & 2 & 2 & 2 & 2 & 2 & 1 & 2 & 3 & 2 & 2 & 2 & 2 & 2 & 1 & 2 & 2.1 (0.7)\\ \hline
\rowcolor{myredfor2}
P37 & 2 & 1 & 1 & 2 & 1 & 2 & 3 & 4 & 4 & 4 & 1 & 2 & 3 & 1 & 1 & 4 & 2.2 (1.2)\\ \hline
\rowcolor{mycyanfor3}
P39 & 3 & 2 & 2 & 1 & 1 & 4 & 2 & 3 & 4 & 4 & 4 & 4 & 4 & 4 & 3 & 4 & 3.1 (1.1)\\ \hline
\rowcolor{myredfor1}
P41 & 1 & 1 & 1 & 1 & 1 & 1 & 1 & 1 & 1 & 1 & 1 & 1 & 1 & 1 & 1 & 1 & 1.0 (0.0)\\ \hline
\rowcolor{myredfor2}
P42 & 3 & 1 & 4 & 4 & 1 & 1 & 2 & 4 & 4 & 2 & 2 & 1 & 4 & 2 & 2 & 4 & 2.6 (1.3)\\ \hline
\rowcolor{myredfor2}
P43 & 2 & 2 & 2 & 2 & 2 & 2 & 2 & 2 & 2 & 2 & 2 & 2 & 2 & 2 & 2 & 2 & 2.0 (0.0)\\ \hline
\rowcolor{myredfor2}
P46 & 3 & 4 & 4 & 5 & 4 & 4 & 4 & 3 & 3 & 2 & 2 & 2 & 1 & 1 & 1 & 4 & 2.9 (1.3)\\ \hline

    \end{tabular}}
    \caption{The sharing tendencies of participants for each emoji in Likert Scale from 1 - ``Very Unlikely'' to 5 - ``Very Likely''. Standard deviations for the averages
are given in the parentheses.}
    \label{tab:participant_tendencies_for_emojis_full_table}
\end{table}

\section{Game Data}\label{sec:all_simulation_data}

Complete data for each participant in the game is provided in Table \ref{tab:participant_all_data_part4}. In order to improve its readability, this table is spread into following four pages.

\setlength{\tabcolsep}{1pt}
\renewcommand{\arraystretch}{1}

\begin{landscape}
\begin{table}[h!]
    \centering
    \scalebox{0.62}{
\begin{tabular}{|c|cc|cc|cc|cc|cc|cc|cc|cc|cc|cc|cc|cc|cc|cc|cc|cc|}
        \hline
    \multirow{2}{*}{}
    & \multicolumn{32}{c|}{\textbf{Scenarios}} \\
        \cline{2-33}
          \rotatebox[origin=c]{85}{\textbf{Participant}} &  \rotatebox[origin=c|]{85}{$\mathbf{S_1}$} & \rotatebox[origin=c]{85}{\textbf{S(5)  \& S(5) \& S(5)}} &  \rotatebox[origin=c|]{85}{$\mathbf{S_2}$} & \rotatebox[origin=c]{85}{\textbf{S(10)  \& N(10) \& S(10)}} &  \rotatebox[origin=c]{85}{$\mathbf{S_3}$} & \rotatebox[origin=c]{85}{\textbf{N(5)  \& S(5) \& N(5)}}&  \rotatebox[origin=c]{85}{$\mathbf{S_4}$} & \rotatebox[origin=c]{85}{\textbf{N(5)  \& N(10) \& N(10)}}&  \rotatebox[origin=c]{85}{$\mathbf{S_5}$} & \rotatebox[origin=c]{85}{\textbf{N(10)  \& N(10) \& N(15)}}&  \rotatebox[origin=c]{85}{$\mathbf{S_6}$} & \rotatebox[origin=c]{85}{\textbf{N(10)  \& S(15) \& N(5)}}&  \rotatebox[origin=c]{85}{$\mathbf{S_7}$} & \rotatebox[origin=c]{85}{\textbf{S(5)  \& N(5) \& S(10)}}&  \rotatebox[origin=c]{85}{$\mathbf{S_8}$} & \rotatebox[origin=c]{85}{\textbf{S(5)  \& S(5) \& S(10)}}&  \rotatebox[origin=c]{85}{$\mathbf{S_9}$} & \rotatebox[origin=c]{85}{\textbf{S(5)  \& S(5) \& S(5)}}&  \rotatebox[origin=c]{85}{$\mathbf{S_{10}}$} & \rotatebox[origin=c]{85}{\textbf{S(5)  \& N(10) \& S(5)}}&  \rotatebox[origin=c]{85}{$\mathbf{S_{11}}$}& \rotatebox[origin=c]{85}{\textbf{N(10)  \& S(5) \& N(5)}} &  \rotatebox[origin=c]{85}{$\mathbf{S_{12}}$}& \rotatebox[origin=c]{85}{\textbf{N(10)  \& N(10) \& N(10)}} &  \rotatebox[origin=c]{85}{$\mathbf{S_{13}}$}& \rotatebox[origin=c]{85}{\textbf{N(5)  \& N(5) \& N(5)}} &  \rotatebox[origin=c]{85}{$\mathbf{S_{14}}$}& \rotatebox[origin=c]{85}{\textbf{N(15)  \& S(10) \& N(5)}} &  \rotatebox[origin=c]{85}{$\mathbf{S_{15}}$} & \rotatebox[origin=c]{85}{\textbf{S(5)  \& N(15) \& S(5)}}&  \rotatebox[origin=c]{85}{$\mathbf{S_{16}}$} & \rotatebox[origin=c]{85}{\textbf{S(10)  \& S(15) \& S(15)}} \\
\hline

    \end{tabular}}
    %\caption{The sharing tendencies of participants for each emoji in Likert Scale from 1 - ``Very Unlikely'' to 5 - ``Very Likely''.}
    \label{tab:participant_all_data_part1}
\end{table}
\end{landscape}

\begin{landscape}
\begin{table}[h]
    \centering
    \scalebox{0.62}{
    \begin{tabular}{|c|cc|cc|cc|cc|cc|cc|cc|cc|cc|cc|cc|cc|cc|cc|cc|cc|}
        \hline
    \multirow{2}{*}{}
    & \multicolumn{32}{c|}{\textbf{Scenarios}} \\
        \cline{2-33}
          \rotatebox[origin=c]{85}{\textbf{Participant}} &  \rotatebox[origin=c|]{85}{$\mathbf{S_1}$} & \rotatebox[origin=c]{85}{\textbf{S(5)  \& S(5) \& S(5)}} &  \rotatebox[origin=c|]{85}{$\mathbf{S_2}$} & \rotatebox[origin=c]{85}{\textbf{S(10)  \& N(10) \& S(10)}} &  \rotatebox[origin=c]{85}{$\mathbf{S_3}$} & \rotatebox[origin=c]{85}{\textbf{N(5)  \& S(5) \& N(5)}}&  \rotatebox[origin=c]{85}{$\mathbf{S_4}$} & \rotatebox[origin=c]{85}{\textbf{N(5)  \& N(10) \& N(10)}}&  \rotatebox[origin=c]{85}{$\mathbf{S_5}$} & \rotatebox[origin=c]{85}{\textbf{N(10)  \& N(10) \& N(15)}}&  \rotatebox[origin=c]{85}{$\mathbf{S_6}$} & \rotatebox[origin=c]{85}{\textbf{N(10)  \& S(15) \& N(5)}}&  \rotatebox[origin=c]{85}{$\mathbf{S_7}$} & \rotatebox[origin=c]{85}{\textbf{S(5)  \& N(5) \& S(10)}}&  \rotatebox[origin=c]{85}{$\mathbf{S_8}$} & \rotatebox[origin=c]{85}{\textbf{S(5)  \& S(5) \& S(10)}}&  \rotatebox[origin=c]{85}{$\mathbf{S_9}$} & \rotatebox[origin=c]{85}{\textbf{S(5)  \& S(5) \& S(5)}}&  \rotatebox[origin=c]{85}{$\mathbf{S_{10}}$} & \rotatebox[origin=c]{85}{\textbf{S(5)  \& N(10) \& S(5)}}&  \rotatebox[origin=c]{85}{$\mathbf{S_{11}}$}& \rotatebox[origin=c]{85}{\textbf{N(10)  \& S(5) \& N(5)}} &  \rotatebox[origin=c]{85}{$\mathbf{S_{12}}$}& \rotatebox[origin=c]{85}{\textbf{N(10)  \& N(10) \& N(10)}} &  \rotatebox[origin=c]{85}{$\mathbf{S_{13}}$}& \rotatebox[origin=c]{85}{\textbf{N(5)  \& N(5) \& N(5)}} &  \rotatebox[origin=c]{85}{$\mathbf{S_{14}}$}& \rotatebox[origin=c]{85}{\textbf{N(15)  \& S(10) \& N(5)}} &  \rotatebox[origin=c]{85}{$\mathbf{S_{15}}$} & \rotatebox[origin=c]{85}{\textbf{S(5)  \& N(15) \& S(5)}}&  \rotatebox[origin=c]{85}{$\mathbf{S_{16}}$} & \rotatebox[origin=c]{85}{\textbf{S(10)  \& S(15) \& S(15)}} \\
\hline

\end{tabular}}
\label{tab:participant_all_data_part2}
\end{table}
\end{landscape}

\begin{landscape}
\begin{table}[h]
    \centering
    \scalebox{0.62}{
    \begin{tabular}{|c|cc|cc|cc|cc|cc|cc|cc|cc|cc|cc|cc|cc|cc|cc|cc|cc|}
        \hline
    \multirow{2}{*}{}
    & \multicolumn{32}{c|}{\textbf{Scenarios}} \\
        \cline{2-33}
          \rotatebox[origin=c]{85}{\textbf{Participant}} &  \rotatebox[origin=c|]{85}{$\mathbf{S_1}$} & \rotatebox[origin=c]{85}{\textbf{S(5)  \& S(5) \& S(5)}} &  \rotatebox[origin=c|]{85}{$\mathbf{S_2}$} & \rotatebox[origin=c]{85}{\textbf{S(10)  \& N(10) \& S(10)}} &  \rotatebox[origin=c]{85}{$\mathbf{S_3}$} & \rotatebox[origin=c]{85}{\textbf{N(5)  \& S(5) \& N(5)}}&  \rotatebox[origin=c]{85}{$\mathbf{S_4}$} & \rotatebox[origin=c]{85}{\textbf{N(5)  \& N(10) \& N(10)}}&  \rotatebox[origin=c]{85}{$\mathbf{S_5}$} & \rotatebox[origin=c]{85}{\textbf{N(10)  \& N(10) \& N(15)}}&  \rotatebox[origin=c]{85}{$\mathbf{S_6}$} & \rotatebox[origin=c]{85}{\textbf{N(10)  \& S(15) \& N(5)}}&  \rotatebox[origin=c]{85}{$\mathbf{S_7}$} & \rotatebox[origin=c]{85}{\textbf{S(5)  \& N(5) \& S(10)}}&  \rotatebox[origin=c]{85}{$\mathbf{S_8}$} & \rotatebox[origin=c]{85}{\textbf{S(5)  \& S(5) \& S(10)}}&  \rotatebox[origin=c]{85}{$\mathbf{S_9}$} & \rotatebox[origin=c]{85}{\textbf{S(5)  \& S(5) \& S(5)}}&  \rotatebox[origin=c]{85}{$\mathbf{S_{10}}$} & \rotatebox[origin=c]{85}{\textbf{S(5)  \& N(10) \& S(5)}}&  \rotatebox[origin=c]{85}{$\mathbf{S_{11}}$}& \rotatebox[origin=c]{85}{\textbf{N(10)  \& S(5) \& N(5)}} &  \rotatebox[origin=c]{85}{$\mathbf{S_{12}}$}& \rotatebox[origin=c]{85}{\textbf{N(10)  \& N(10) \& N(10)}} &  \rotatebox[origin=c]{85}{$\mathbf{S_{13}}$}& \rotatebox[origin=c]{85}{\textbf{N(5)  \& N(5) \& N(5)}} &  \rotatebox[origin=c]{85}{$\mathbf{S_{14}}$}& \rotatebox[origin=c]{85}{\textbf{N(15)  \& S(10) \& N(5)}} &  \rotatebox[origin=c]{85}{$\mathbf{S_{15}}$} & \rotatebox[origin=c]{85}{\textbf{S(5)  \& N(15) \& S(5)}}&  \rotatebox[origin=c]{85}{$\mathbf{S_{16}}$} & \rotatebox[origin=c]{85}{\textbf{S(10)  \& S(15) \& S(15)}} \\
\hline

\end{tabular}}
\label{tab:participant_all_data_part3}
\end{table}
\end{landscape}

\begin{landscape}
\begin{table}[h]
    \centering
    \scalebox{0.62}{
    \begin{tabular}{|c|cc|cc|cc|cc|cc|cc|cc|cc|cc|cc|cc|cc|cc|cc|cc|cc|}
        \hline
    \multirow{2}{*}{}
    & \multicolumn{32}{c|}{\textbf{Scenarios}} \\
        \cline{2-33}
          \rotatebox[origin=c]{85}{\textbf{Participant}} &  \rotatebox[origin=c|]{85}{$\mathbf{S_1}$} & \rotatebox[origin=c]{85}{\textbf{S(5)  \& S(5) \& S(5)}} &  \rotatebox[origin=c|]{85}{$\mathbf{S_2}$} & \rotatebox[origin=c]{85}{\textbf{S(10)  \& N(10) \& S(10)}} &  \rotatebox[origin=c]{85}{$\mathbf{S_3}$} & \rotatebox[origin=c]{85}{\textbf{N(5)  \& S(5) \& N(5)}}&  \rotatebox[origin=c]{85}{$\mathbf{S_4}$} & \rotatebox[origin=c]{85}{\textbf{N(5)  \& N(10) \& N(10)}}&  \rotatebox[origin=c]{85}{$\mathbf{S_5}$} & \rotatebox[origin=c]{85}{\textbf{N(10)  \& N(10) \& N(15)}}&  \rotatebox[origin=c]{85}{$\mathbf{S_6}$} & \rotatebox[origin=c]{85}{\textbf{N(10)  \& S(15) \& N(5)}}&  \rotatebox[origin=c]{85}{$\mathbf{S_7}$} & \rotatebox[origin=c]{85}{\textbf{S(5)  \& N(5) \& S(10)}}&  \rotatebox[origin=c]{85}{$\mathbf{S_8}$} & \rotatebox[origin=c]{85}{\textbf{S(5)  \& S(5) \& S(10)}}&  \rotatebox[origin=c]{85}{$\mathbf{S_9}$} & \rotatebox[origin=c]{85}{\textbf{S(5)  \& S(5) \& S(5)}}&  \rotatebox[origin=c]{85}{$\mathbf{S_{10}}$} & \rotatebox[origin=c]{85}{\textbf{S(5)  \& N(10) \& S(5)}}&  \rotatebox[origin=c]{85}{$\mathbf{S_{11}}$}& \rotatebox[origin=c]{85}{\textbf{N(10)  \& S(5) \& N(5)}} &  \rotatebox[origin=c]{85}{$\mathbf{S_{12}}$}& \rotatebox[origin=c]{85}{\textbf{N(10)  \& N(10) \& N(10)}} &  \rotatebox[origin=c]{85}{$\mathbf{S_{13}}$}& \rotatebox[origin=c]{85}{\textbf{N(5)  \& N(5) \& N(5)}} &  \rotatebox[origin=c]{85}{$\mathbf{S_{14}}$}& \rotatebox[origin=c]{85}{\textbf{N(15)  \& S(10) \& N(5)}} &  \rotatebox[origin=c]{85}{$\mathbf{S_{15}}$} & \rotatebox[origin=c]{85}{\textbf{S(5)  \& N(15) \& S(5)}}&  \rotatebox[origin=c]{85}{$\mathbf{S_{16}}$} & \rotatebox[origin=c]{85}{\textbf{S(10)  \& S(15) \& S(15)}} \\
\hline

\end{tabular}}
\caption{Game data of each participant who passed the quiz and completed the study. Fixed decisions and bids for the agents are included in the header of each scenario. Data of each participant are provided in group of 4 rows, where the first row denotes the initial tendencies to share in Likert scale (From $L_1$ - ``Very Unlikely'' to $L_5$ - ``Very Likely''). Initial and final privacy choice of the participant is given in the second row along with the initial and final bids in parentheses. Third row provides the outcomes of first and second auctions. Finally, last row in each group labels the behaviors in the scenarios as described in Figure \ref{fig:all_behaviors_p2}.}
\label{tab:participant_all_data_part4}
\end{table}
\end{landscape}

\section{Training Material}\label{sec:training_material}

\subsection{Slides}

\mypara{\textbf{Page 1:}}

In the following pages, we will show you examples of scenarios that you will be presented as part of this research in order to familiarize you with the experiment.

\textcolor{red}{PLEASE PAY ATTENTION!}

\vspace{5mm}
\mypara{\textbf{Page 2:}}

\textcolor{red}{PLEASE PAY ATTENTION:} We will first explain how the experiment works, and then we will ask you questions to see if you have understood various aspects of the experiment.

\textcolor{red}{PLEASE BE AWARE:} If you make errors during the training, you will have a chance to repeat it. However, in order to participate in the remainder of the study and receive payment you will have to pass the quiz in the training.

\vspace{5mm}
\mypara{\textbf{Page 3:}} 
\begin{figure}[H]    \includegraphics[scale=0.4, left]{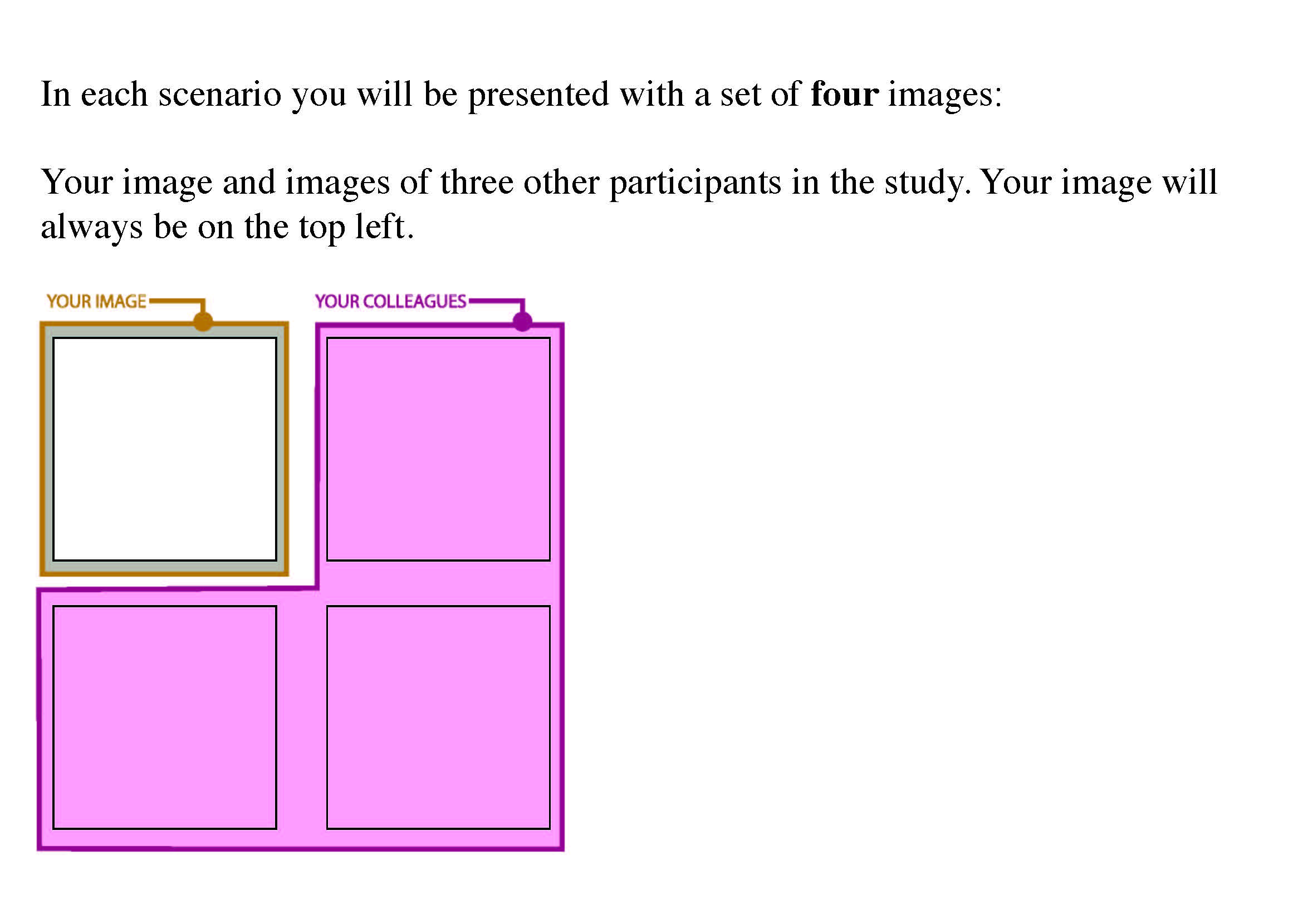}
    \label{fig:ts_page3}
\end{figure}

\clearpage
\mypara{\textbf{Page 4:}}

\textbf{FOR EACH SCENARIO WE ARE ASKING YOU TO IMAGINE THIS SITUATION:}

You (yourself and your colleagues) have taken a group picture and need to decide if you want it to be posted online. To reach an outcome, we are asking you to assign a preference (share or not share) and a value (in credits) to support your decision. 

After you all have made your individual decision, you will be shown each other’s sharing preferences and whether or not your preference will be the favorable outcome.

Look at your colleagues’ preferences compared to yours. You can now choose to change or keep your decision and the associated credits based on that.

\textcolor{red}{PLEASE BE AWARE that the decision at the second round will be final whether your choice is favorable or not.}

\mypara{\textbf{Page 5:}}

Your task is to decide whether to \textbf{SHARE} or \textbf{NOT SHARE} all the images in this scenario. You have an amount of credits (budget) that you can use to reinforce your decision and enter an amount to do so.  The credit(s) you can use in the scenarios come from your budget. For this experiment, we give 50 credits as the initial budget that you can spend from, and for each scenario you will get an additional 10.

The credit(s) you place on an outcome reflects how much the decision matters to you. You will be able to place an amount between 1 and 20 for a decision, as long as you have enough budget to do so.

Enter the amount you want to place and then select your sharing decision. Once you are certain of your choices you can submit.

\mypara{\textbf{Page 6:}}
\begin{figure}[H]    \includegraphics[scale=0.36, left]{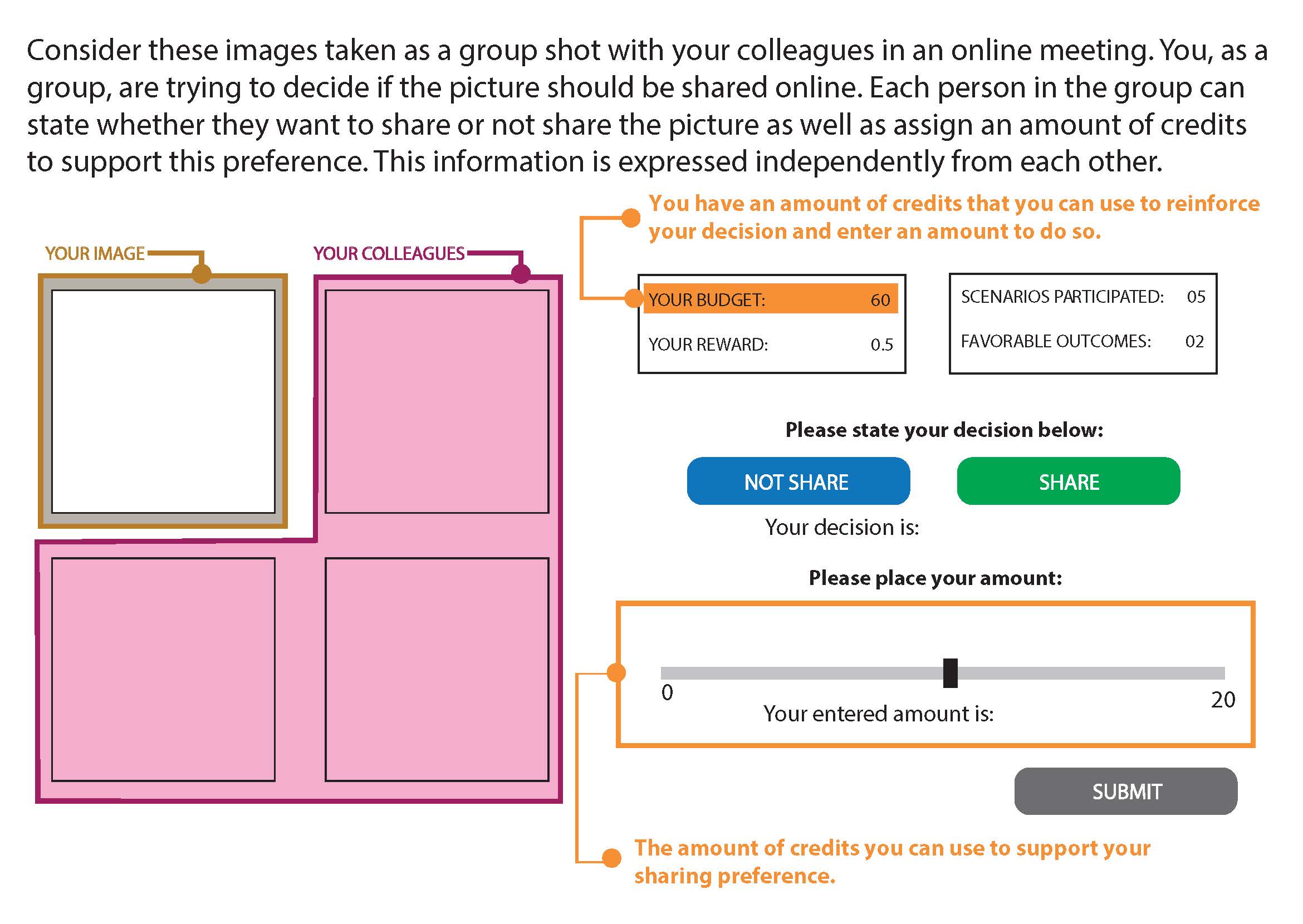}
    \label{fig:ts_page6}
\end{figure}

\clearpage

\mypara{\textbf{Page 7:}}
\begin{figure}[H]    \includegraphics[scale=0.36, left]{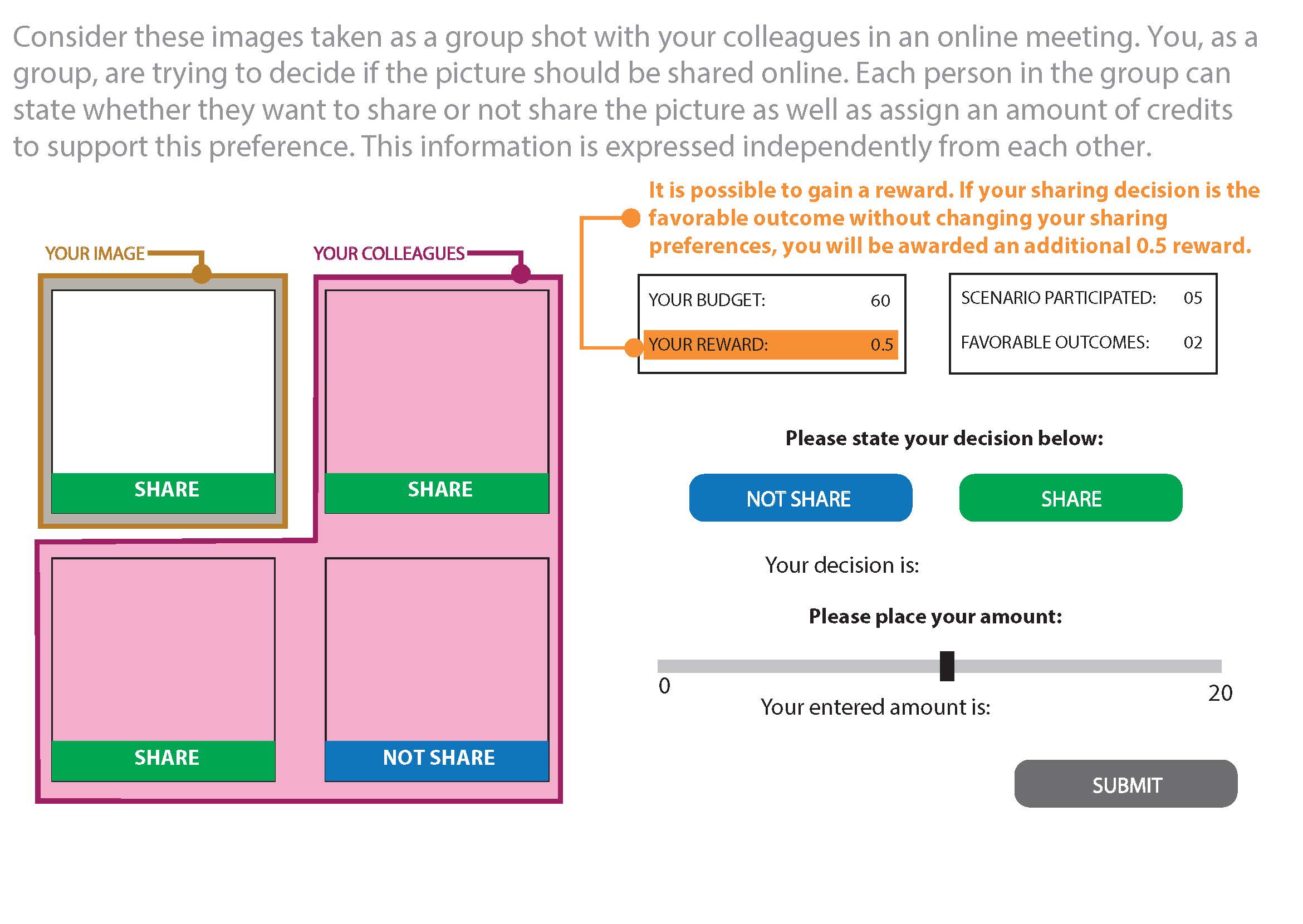}
    \label{fig:ts_page7}
\end{figure}

\mypara{\textbf{Page 8:}}
\begin{figure}[H]    \includegraphics[scale=0.36, left]{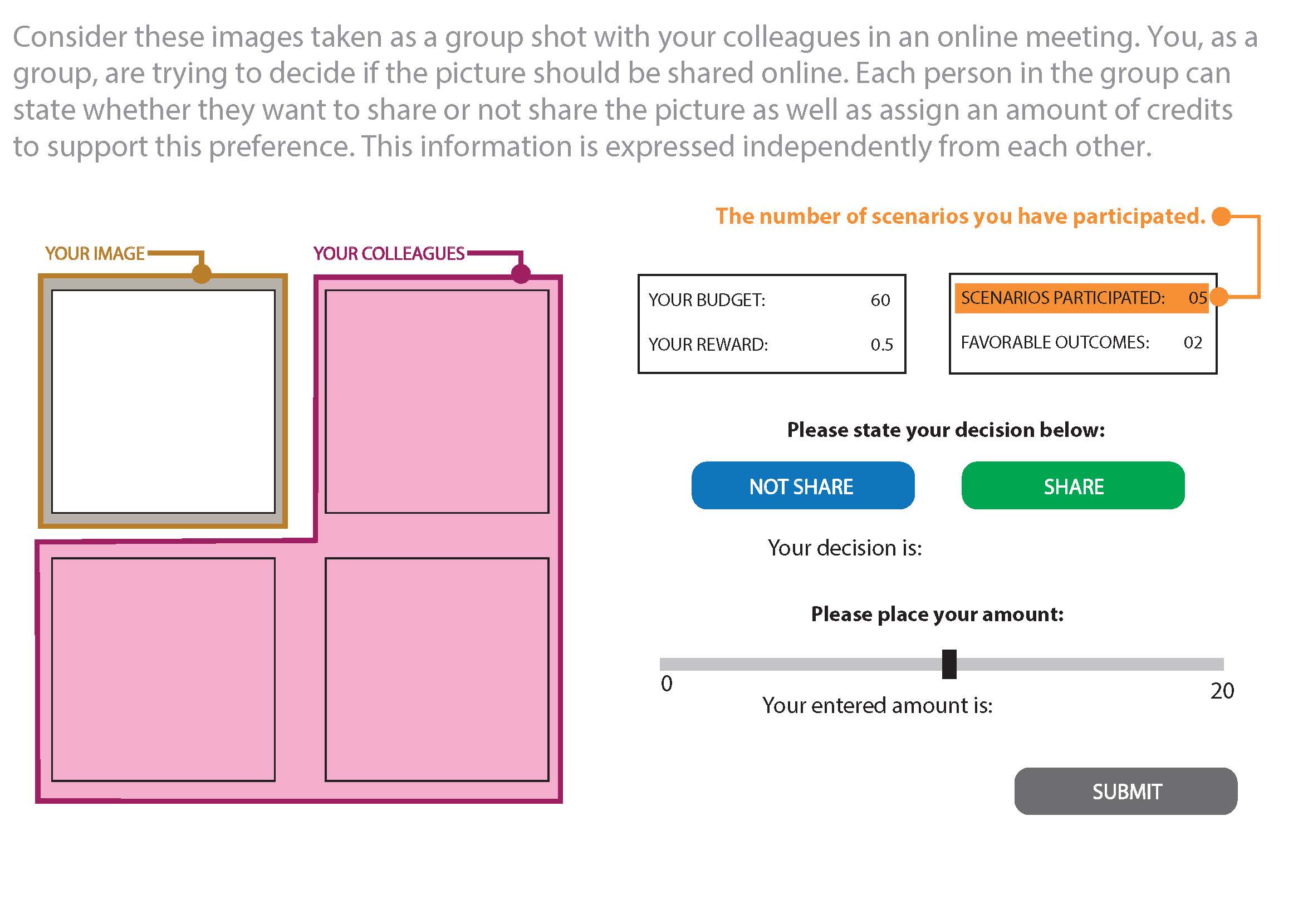}
    \label{fig:ts_page8}
\end{figure}

\mypara{\textbf{Page 9:}}
\begin{figure}[H]    \includegraphics[scale=0.36, left]{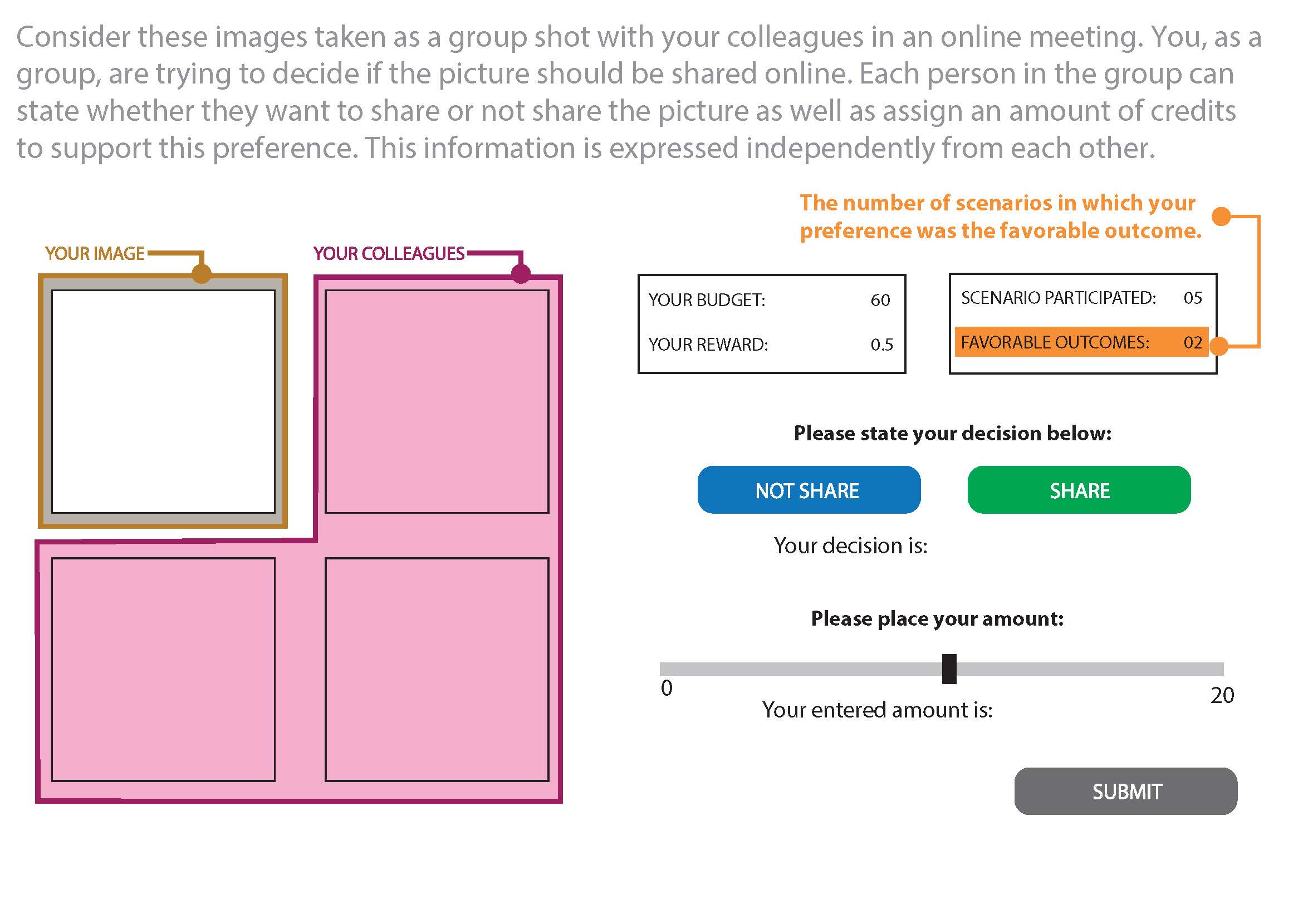}
    \label{fig:ts_page9}
\end{figure}

\mypara{\textbf{Page 10:}}
\begin{figure}[H]    \includegraphics[scale=0.36, left]{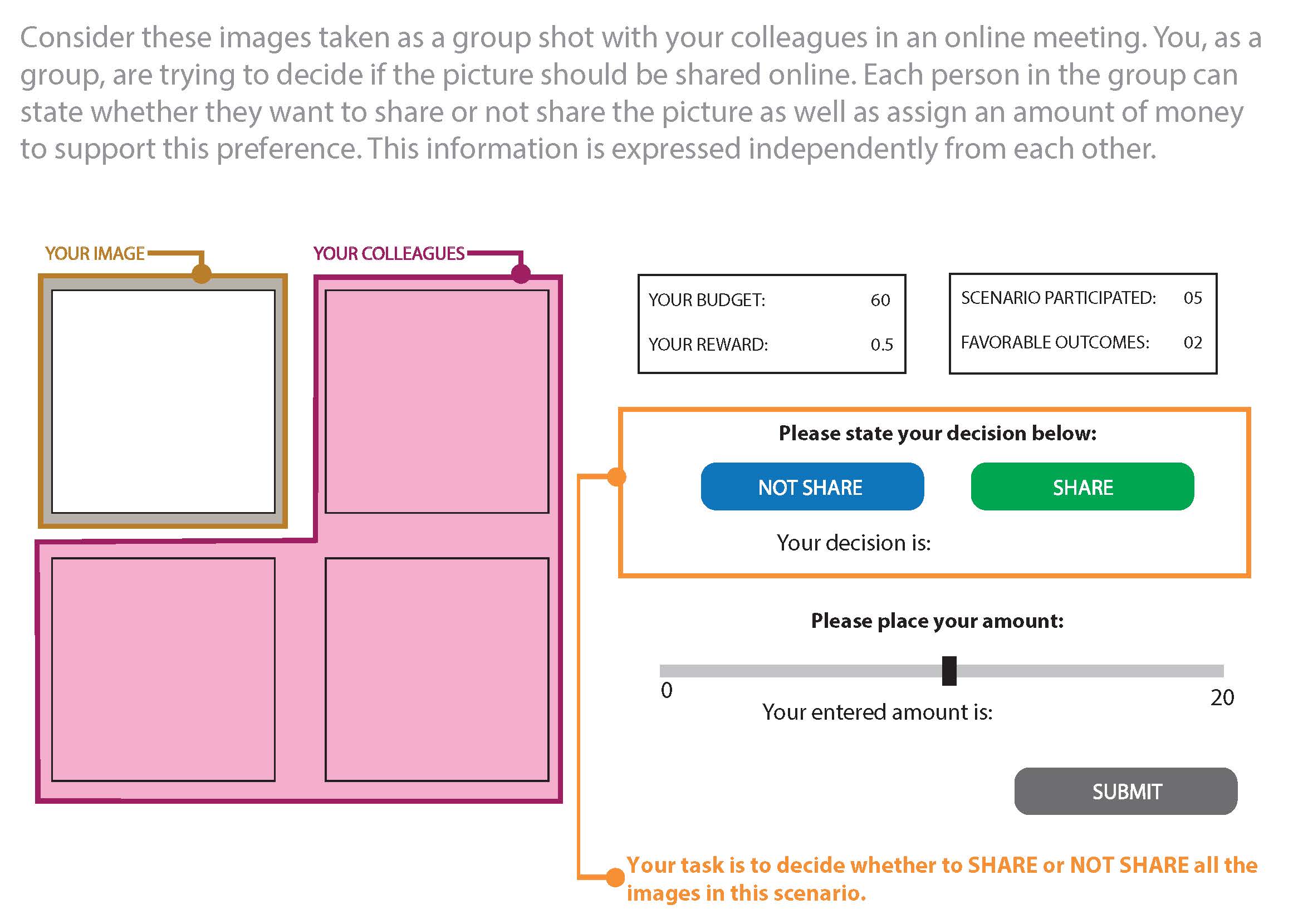}
    \label{fig:ts_page10}
\end{figure}

\mypara{\textbf{Page 11:}}
\begin{figure}[H]    \includegraphics[scale=0.36, left]{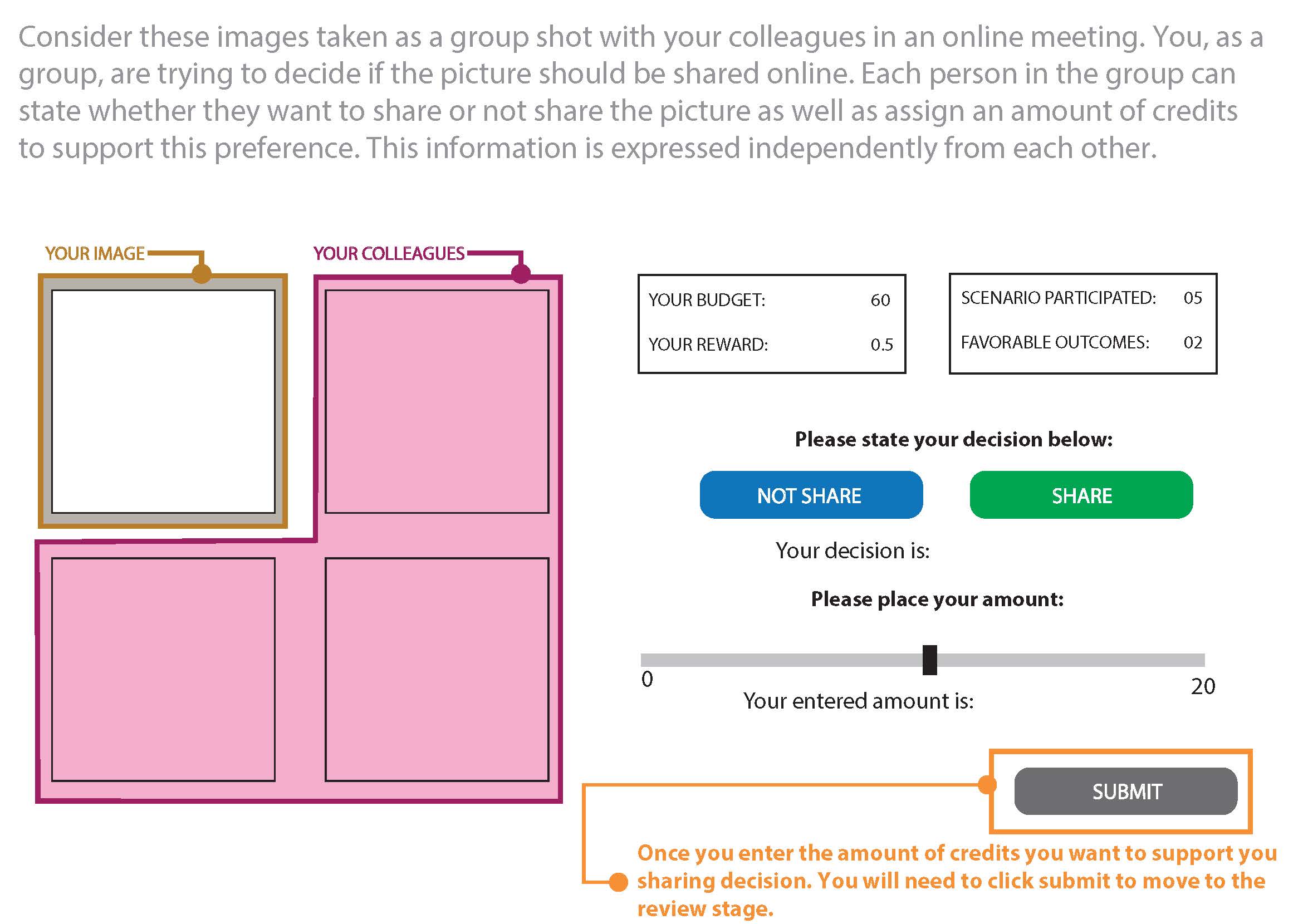}
    \label{fig:ts_page11}
\end{figure}

\vspace{-10mm}
\mypara{\textbf{Page 12:}}

You will be taken to another page where all the other participants’ sharing preferences will be shown. In this stage you can either decide to keep or change your selection. At this time, the amounts for \textbf{SHARE} and \textbf{NOT SHARE} will be added individually. The decision with the highest total will be the outcome.

\mypara{\textbf{Page 13:}}

Please be aware that if your decision does not match the final outcome (i.e. your sharing decision is not the favorable outcome), you will not be charged. If it does (i.e. your sharing decision is the favorable outcome), you will be charged. Your sharing decision may be the favorable outcome because you have placed more credits than all the other participants for an outcome or that it could be favorable as part of the consensus.

For example: \\
\vspace{-3mm}

You - 5 Not Share \\
\indent Participant \#2 - 5 Share \\
\indent Participant \#3 - 10 Share \\
\indent Participant \#4 - 5 Not Share \\

\vspace{-3mm}
You will not be charged. \\

\vspace{-3mm}
You - 5 Share \\
\indent Participant \#2 - 5 Share \\
\indent Participant \#3 - 10 Share \\
\indent Participant \#4 - 5 Not Share \\

\vspace{-3mm}
You will be charged.

\clearpage
\mypara{\textbf{Page 14:}}

\textbf{How much will you be charged?}

If your sharing decision is the favorable outcome but your amount did not affect the outcome of the scenario (e.g. the outcome would have been the same without you), you will only be charged the amount you have specified.

For example: \\

You - 5 Not Share \\
\indent Participant \#2 - 5 Share \\
\indent Participant \#3 - 5 Share \\
\indent Participant \#4 - 15 Not Share \\

You will be charged 5. \\

However, if your sharing decision is the favorable outcome and your amount did affect the outcome an extra 5 will be deducted from your budget. This is what we call a tax.

For example: \\

You - 15 Share \\
\indent Participant \#2 - 5 Share \\
\indent Participant \#3 - 5 Share \\
\indent Participant \#4 - 15 Not Share \\

Here you will be charged the amount you entered (10) and also an extra amount of 5 (tax) since without your amount the outcome would have been different. Hence, you will be charged 15 in total.

\mypara{\textbf{Page 15:}}
\textbf{How much will you gain?}

When you finish this study, you will be rewarded a minimum of €5. It is possible to gain an extra payment. If your sharing decision is the favorable outcome without changing your sharing preferences you will be awarded an additional 0.5 reward. You still get the same reward if you only change the amount you entered without changing your sharing preferences and your sharing decision is the favorable outcome.

For example: \\

\textit{First Round:} \\

You - 5 Share \\
\indent Participant \#2 - 5 Share \\
\indent Participant \#3 - 10 Share \\
\indent Participant \#4 - 5 Not  Share \\

\textit{Second Round:} \\

You - 5 Share \\
\indent Participant \#2 - 5 Share \\
\indent Participant \#3 - 10 Share \\
\indent Participant \#4 - 5 Not  Share \\

Here you will get a reward of 0.5. \\

The amount of credits left in your budget and in the extra reward will be converted to additional payment. For each 10 you save at the end of the experiment from your budget, you will get an additional 0.1 reward (e.g 10 -> 0.1; 15 -> 0.1; 23 -> 0.2). The total reward at the end of the experiment will be converted to lottery tickets to win an Amazon voucher worth €20.

\subsection{Quiz}

As mentioned before, questions are displayed in a sample scenario with a sample group photo including the participant, as in the case of actual experiment. Agents in the given scenario are named as Participant \#2, Participant \#3 and Participant \#4 so that the actual participant can consider them as peers.

Regardless its type (interactive task or multiple choice), all possible phrases of feedback to the participant for a question is listed here as well. For the first two questions, more than one phrase can be displayed to the participant, if their actions in the given task include more than one mistake. Feedback that is associated with the correct option (or the correct action in the interactive questions) is marked with `*', as well as the correct option.  

%%%% Q1 %%%%
\mypara{\textbf{Question 1:}}

\underline{Your training task is:} \\

To pick a preference of \textbf{NOT SHARE}

and to enter an amount of money \textbf{higher than or equal to 11.}

\mypara{\textbf{Feedback 1.1:}} You have not entered a decision!

\mypara{\textbf{Feedback 1.2:}} You have not entered an amount of credits!

\mypara{\textbf{Feedback 1.3:}} You have entered ``SHARE'' instead of ``NOT SHARE'' as your decision. The instructions stated to enter ``NOT SHARE''.

\mypara{\textbf{Feedback 1.4:}} You have not entered the correct amount of credits that was specified in the instructions.

\mypara{\textbf{Feedback 1.5*:}} Correct! You have correctly entered ``NOT SHARE'' as your decision and correctly entered the amount of credits.

%%%% Q2 %%%%
\mypara{\textbf{Question 2:}}

\underline{Your training task is:} \\

To pick a preference of \textbf{SHARE}

and to enter an amount of money \textbf{lower than or equal to 7.}

\clearpage
\mypara{\textbf{Feedback 2.1:}} You have not entered a decision!

\mypara{\textbf{Feedback 2.2:}} You have not entered an amount of credits!

\mypara{\textbf{Feedback 2.3:}} You have entered ``NOT SHARE'' instead of ``SHARE'' as your decision. The instructions stated to enter ``SHARE''.

\mypara{\textbf{Feedback 2.4:}} You have not entered the correct amount of credits that was specified in the instructions.

\mypara{\textbf{Feedback 2.5*:}} Correct! You have correctly entered ``SHARE'' as your decision and correctly entered the amount of credits.

%%%% Q3 %%%%
\mypara{\textbf{Question 3:}}

\underline{Your training task is:} 

To specify what the outcome of the scenario is given all the participants sharing preferences and amount of credits entered.

\textbf{Your Choices:} \\
\indent \indent Sharing reference: 	``NOT SHARE'' \\
\indent \indent Amount:		13 

\textbf{Participant \#2:} \\ 
\indent \indent Sharing preference:	``SHARE'' \\
\indent \indent Amount: 		10

\textbf{Participant \#3:} \\
\indent \indent Sharing preference:	``SHARE'' \\
\indent \indent Amount: 		10

\textbf{Participant \#4:} \\
\indent \indent Sharing preference:	``NOT SHARE'' \\
\indent \indent Amount: 		10 \\ 

What will be the outcome of the scenario? \\
\indent \textbf{a)} SHARE \indent \indent \textbf{b*)} NOT SHARE 

\mypara{\textbf{Feedback 3.1:}} The outcome of the scenario should have been ``NOT SHARE'' because the amount of credits entered by the participants that selected ``NOT SHARE'' is greater than the amount of credits entered by participants selecting ``SHARE''.

\mypara{\textbf{Feedback 3.2*:}} Correct! You have correctly selected the outcome of the scenario. The amount of credits entered by the participants that selected ``NOT SHARE'' is greater than the amount of credits entered by participants selecting ``SHARE''.

%%%% Q4 %%%%
\mypara{\textbf{Question 4:}}

\underline{Your training task is:} 

To specify what the outcome of the scenario is given all the participants sharing preferences and amount of credits entered.

\clearpage
\textbf{Your Choices:} \\
\indent \indent Sharing reference: 	``SHARE'' \\
\indent \indent Amount:		5 

\textbf{Participant \#2:} \\ 
\indent \indent Sharing preference:	``SHARE'' \\
\indent \indent Amount: 		5

\textbf{Participant \#3:} \\
\indent \indent Sharing preference:	``SHARE'' \\
\indent \indent Amount: 		5

\textbf{Participant \#4:} \\
\indent \indent Sharing preference:	``NOT SHARE'' \\
\indent \indent Amount: 		20  

What will be the outcome of the scenario? \\
\indent \textbf{a)} SHARE \indent \indent \textbf{b*)} NOT SHARE 

\mypara{\textbf{Feedback 4.1:}} The outcome of the scenario should have been ``NOT SHARE'' because the amount of credits entered by the participants that selected ``NOT SHARE'' is greater than the amount of credits entered by participants selecting ``SHARE''.

\mypara{\textbf{Feedback 4.2*:}} Correct! You have correctly selected the outcome of the scenario. The amount of credits entered by the participants that selected ``NOT SHARE'' is greater than the amount of credits entered by participants selecting ``SHARE''.

%%%% Q5 %%%%
\mypara{\textbf{Question 5:}}

\underline{Your training task is:}

To specify the amount of money that you will pay in this scenario. 

In the actual experiment, the amount you enter will be deducted from your available budget in case your sharing decision is the favorable outcome.

\textbf{REMEMBER:  If your decision changes the outcome, you will be charged an additional 5 as tax.}

\textbf{Your Choices:} \\
\indent \indent Sharing reference: 	``NOT SHARE'' \\
\indent \indent Amount:		15 

\textbf{Participant \#2:} \\ 
\indent \indent Sharing preference:	``SHARE'' \\
\indent \indent Amount: 		10

\textbf{Participant \#3:} \\
\indent \indent Sharing preference:	``SHARE'' \\
\indent \indent Amount: 		10

\textbf{Participant \#4:} \\
\indent \indent Sharing preference:	``NOT SHARE'' \\
\indent \indent Amount: 		10  

How much would you pay in this situation? \\
\indent \textbf{a)} 0 \indent \indent \textbf{b)} 10
\indent \indent \textbf{c)} 15
\indent \indent \textbf{d*)} 20

\clearpage
\mypara{\textbf{Feedback 5.1:}} Selecting 0 is never an option, since the scenario's outcome aligned with your sharing preference, you will at least always pay the amount of credits you have specified.

\mypara{\textbf{Feedback 5.2:}} This amount is not in line with the amount that you are supposedly entered (Participant \#1). Please make sure you understand how to participate in this study prior to taking part.

\mypara{\textbf{Feedback 5.3:}} The amount you entered affects the decision of the outcome. This means that without your amount of credits the opposite preference would have prevailed. In this case you will pay an additional 5 from your budget. Please make sure you understand how to participate in this study prior to taking part.

\mypara{\textbf{Feedback 5.4*:}} Correct! You are correct, you will pay 20 from your budget. Since the amount you entered affected the outcome of the scenario you will be charged an additional amount (5) in this case.

%%%% Q6 %%%%
\mypara{\textbf{Question 6:}}

\underline{Your training task is:} 

To specify the amount of money that you will pay in this scenario. 

In the actual experiment, the amount you enter will be deducted from your available budget in case your sharing decision is the favorable outcome.

\textbf{REMEMBER:  If your decision changes the outcome, you will be charged an additional 5 as tax.}

\textbf{Your Choices:} \\
\indent \indent Sharing reference: 	``NOT SHARE'' \\
\indent \indent Amount:		9 

\textbf{Participant \#2:} \\ 
\indent \indent Sharing preference:	``SHARE'' \\
\indent \indent Amount: 		3

\textbf{Participant \#3:} \\
\indent \indent Sharing preference:	``SHARE'' \\
\indent \indent Amount: 		5

\textbf{Participant \#4:} \\
\indent \indent Sharing preference:	``NOT SHARE'' \\
\indent \indent Amount: 		10  

How much would you pay in this situation? \\
\indent \textbf{a)} 0 \indent \indent \textbf{b*)} 9
\indent \indent \textbf{c)} 14
\indent \indent \textbf{d)} 19

\mypara{\textbf{Feedback 6.1:}} Selecting 0 is never an option, since the scenario's outcome aligned with your sharing preference, you will at least always pay the amount of credits you have specified.

\mypara{\textbf{Feedback 6.2*:}} Correct! You are correct, you will pay 9 from your budget. Since the credits you placed for your sharing decision did not affect the outcome of the scenario you will not be charged an additional amount in this case.

\mypara{\textbf{Feedback 6.3:}} You will not pay 14 for this scenario. The amount of credits you entered did not affect the outcome hence you will not be charged the extra 5 credits. You would only have paid the additional credits if participant \#4 had entered an amount lower than 8. In that case since your amount would have been higher on its own compared to the SHARE preference amount, you would have been taxed.

\clearpage
\mypara{\textbf{Feedback 6.4:}} This amount is not in line with the amount that you are supposedly entered (Participant \#1). Please make sure you understand how to participate in this study prior to taking part.

%%%% Q7 %%%%
\mypara{\textbf{Question 7:}}

\underline{Your training task is:} 

To specify the amount of money that you will pay in this scenario. 

In the actual experiment, the amount you enter will be deducted from your available budget in case your sharing decision is the favorable outcome.

\textbf{REMEMBER:  If your decision changes the outcome, you will be charged an additional 5 as tax.}

\textbf{Your Choices:} \\
\indent \indent Sharing reference: 	``SHARE'' \\
\indent \indent Amount:		8 

\textbf{Participant \#2:} \\ 
\indent \indent Sharing preference:	``SHARE'' \\
\indent \indent Amount: 		9

\textbf{Participant \#3:} \\
\indent \indent Sharing preference:	``NOT SHARE'' \\
\indent \indent Amount: 		14

\textbf{Participant \#4:} \\
\indent \indent Sharing preference:	``NOT SHARE'' \\
\indent \indent Amount: 		7 

How much would you pay in this situation? \\
\indent \textbf{a*)} 0 \indent \indent \textbf{b)} 8
\indent \indent \textbf{c)} 13
\indent \indent \textbf{d)} 20

\mypara{\textbf{Feedback 7.1*:}} Correct! You will not pay anything for this scenario since even though you entered 8, the decision you selected is not the outcome of the scenario.

\mypara{\textbf{Feedback 7.2:}} You will not pay anything for this scenario since even though you entered 8, the decision you selected is not the outcome of the scenario.

\mypara{\textbf{Feedback 7.3:}} You will not pay anything for this scenario. REMEMBER: only if the scenario outcome aligns with your preference you pay the amount of credits entered and you only pay the extra 5 credits if the amount of credits you enter affects the outcome. Neither of these happened in this scenario. Please make sure you understand how to participate in this study prior to taking part.

\mypara{\textbf{Feedback 7.4:}} You have not entered this amount for the decision, therefore you are not paying the amount you have selected.

%%%% Q8 %%%%
\mypara{\textbf{Question 8:}}

\underline{Your training task is:}  

To specify the amount of money that you will pay in this scenario. 

In the actual experiment, the amount you enter will be deducted from your available budget in case your sharing decision is the favorable outcome.

\textbf{REMEMBER:  If your decision changes the outcome, you will be charged an additional 5 as tax.}
\clearpage

\textbf{Your Choices:} \\
\indent \indent Sharing reference: 	``NOT SHARE'' \\
\indent \indent Amount:		20 

\textbf{Participant \#2:} \\ 
\indent \indent Sharing preference:	``SHARE'' \\
\indent \indent Amount: 		9

\textbf{Participant \#3:} \\
\indent \indent Sharing preference:	``SHARE'' \\
\indent \indent Amount: 		7

\textbf{Participant \#4:} \\
\indent \indent Sharing preference:	``SHARE'' \\
\indent \indent Amount: 		7  

How much would you pay in this situation? \\
\indent \textbf{a*)} 0 \indent \indent \textbf{b)} 7
\indent \indent \textbf{c)} 9
\indent \indent \textbf{d)} 20

\mypara{\textbf{Feedback 8.1*:}} Correct! You will not pay anything for this scenario since even though you entered 20, the decision you selected is not the outcome of the scenario. 

\mypara{\textbf{Feedback 8.2:}} You have not entered this amount for the decision, therefore you are not paying the amount you have selected.

\mypara{\textbf{Feedback 8.3:}} You will not pay anything for this scenario since even though you entered 20, the decision you selected is not the outcome of the scenario.

%%%% Q9 %%%%
\mypara{\textbf{Question 9:}}

\underline{Your training task is:} 

To select who would get the highest reward given the information.

\textbf{REMEMBER: If your sharing decision is the favorable outcome you earn a reward.}

\textbf{The total reward at the end of the experiment will be converted to lottery tickets to win an Amazon  voucher worth €20.}

\begin{table}[H]
    \centering
    \scalebox{0.9}{\begin{tabular}{|c|c|c|c|}
    \hline
    \textbf{Participant}& \textbf{Scenarios Participated} & \textbf{Favorable Outcomes} & \textbf{Budget} \\ \hline

    You & 16 & 13 & 20 \\ \hline

    Participant \#2 & 16 & 8 & 20 \\ \hline

    Participant \#3 & 16 & 4 & 20 \\ \hline

    Participant \#4 & 16 & 11 & 20 \\ \hline
    \end{tabular}}\label{tab:q9}
\end{table}

Who would get the highest reward? \\
\indent \textbf{a*)} You \indent \indent \textbf{b)} Participant \#2
\indent \indent \textbf{c)} Participant \#3
\indent \indent \textbf{d)} Participant \#4

\mypara{\textbf{Feedback 9.1*:}} Correct! This is correct. Everyone saved the same amount of budget, but you had more favorable outcomes than the others. Therefore in this scenario you would get the highest reward. 

\mypara{\textbf{Feedback 9.2:}} This is incorrect. Everyone saved the same amount of budget, but you had more favorable outcomes than the others. Therefore in this scenario you would get the highest reward, not the other participants.

\clearpage
%%%% Q10 %%%%
\mypara{\textbf{Question 10:}}

\underline{Your training task is:} 

To select who would get the highest reward given the information.

\textbf{REMEMBER: If your sharing decision is the favorable outcome you earn a reward.}

\textbf{The total reward at the end of the experiment will be converted to lottery tickets to win an Amazon  voucher worth €20.}

\begin{table}[H]
    \centering
    \scalebox{0.9}{
    \begin{tabular}{|c|c|c|c|}
    \hline
    \textbf{Participant}& \textbf{Scenarios Participated} & \textbf{Favorable Outcomes} & \textbf{Budget} \\ \hline

    You & 16 & 9 & 22 \\ \hline

    Participant \#2 & 16 & 9 & 43 \\ \hline

    Participant \#3 & 16 & 9 & 37 \\ \hline

    Participant \#4 & 16 & 9 & 42 \\ \hline
    \end{tabular}}\label{tab:q10}
\end{table}

Who would get the highest reward? \\
\indent \textbf{a)} You \indent \indent \textbf{b*)} Participant \#2
\indent \indent \textbf{c)} Participant \#3
\indent \indent \textbf{d)} Participant \#4

\mypara{\textbf{Feedback 10.1:}} This is incorrect. While everyone had the same number of favorable outcomes, Participant \#2 has saved the most budget, therefore gets the highest reward.

\mypara{\textbf{Feedback 10.2*:}} Correct! This is correct. While everyone had the same number of favorable outcomes, Participant \#2 has saved the most budget, therefore gets the highest reward.

\end{document}